\documentclass[english,a4paper,prx,twocolumn,amsmath,amssymb,superscriptaddress,longbibliography]{revtex4-1}
\usepackage{bm}
\usepackage{graphicx,color}
\usepackage{soul}
\usepackage[textsize=small]{todonotes}
\usepackage[colorlinks,linkcolor=magenta,citecolor=magenta]{hyperref}

\newcommand{\dd}{\mathrm{d}}

\newcommand{\pd}[2]{\frac{\partial #1}{\partial #2}}

\newcommand{\mean}[1]{\langle #1 \rangle}
\newcommand{\Int}[1]{\int\dd #1\;}
\newcommand{\IInt}[3]{\int_{#2}^{#3}\dd #1\;}
\newcommand{\OInt}[3]{\oint_{#2}^{#3}\dd #1\;}

\renewcommand{\vec}[1]{\mathbf #1}

\renewcommand{\vec}[1]{\bm{\mathbf{#1}}}

\usepackage{siunitx}

\newcommand{\al}{\alpha}

\newcommand{\eps}{\varepsilon}
\newcommand{\kap}{\kappa}
\newcommand{\lam}{\lambda}
\newcommand{\vhi}{\varphi}
\newcommand{\sig}{\sigma}
\newcommand{\om}{\omega}
\newcommand{\Om}{\Omega}

\newcommand{\id}{\mathbf 1}

\newcommand{\x}{\vec r}

\newcommand{\nois}{\bm\xi}
\newcommand{\kT}{k_\text{B}T}
\newcommand{\rc}{r_\text{C}} 

\newcommand{\js}{\hat{\bm\jmath}} 
\newcommand{\msig}{\bm\sig}
\newcommand{\Fex}{\vec F^\text{ex}}
\newcommand{\keff}{k_\text{eff}}
\newcommand{\dbar}{\dd \hspace*{-0.14em}\bar{}\hspace*{0.1em}}

\AtBeginDocument{%
  \newwrite\bibnotes
  \def\bibnotesext{Notes.bib}
  \immediate\openout\bibnotes=\jobname\bibnotesext
  \immediate\write\bibnotes{@CONTROL{REVTEX41Control}}
  \immediate\write\bibnotes{@CONTROL{apsrev41Control,author="08",editor="1",pages="1",title="0",year="1"}}
  \if@filesw
  \immediate\write\@auxout{\string\citation{apsrev41Control}}%
  \fi
}%

\begin{document}

\title{Thermodynamics of active matter: Tracking dissipation across scales}

\author{Robin Bebon}
\affiliation{Institute for Theoretical Physics IV, University of Stuttgart, 70569 Stuttgart, Germany}
\author{Joshua F. Robinson}
\affiliation{Institut f\"ur Physik, Johannes Gutenberg-Universit\"at Mainz, Staudingerweg 7-9, 55128 Mainz, Germany}
\affiliation{H.\ H.\ Wills Physics Laboratory, University of Bristol, Bristol BS8 1TL, UK}
\author{Thomas Speck}
\affiliation{Institute for Theoretical Physics IV, University of Stuttgart, 70569 Stuttgart, Germany}

\begin{abstract}
  The concept of entropy has been pivotal in the formulation of thermodynamics. For systems driven away from thermal equilibrium, a comparable role is played by entropy production and dissipation. Here we provide a comprehensive picture how local dissipation due to effective chemical events manifests on large scales in active matter. We start from a microscopic model for a single catalytic particle involving explicit solute molecules and show that it undergoes directed motion. Leveraging stochastic thermodynamics, we calculate the average entropy production rate for interacting particles. We then show how the model of active Brownian particles emerges in a certain limit and we determine the entropy production rate on the level of the hydrodynamic equations. Our results augment the model of active Brownian particles with rigorous expressions for the dissipation that cannot be inferred from their equations of motion, and we illustrate consequences for wall aggregation and motility-induced phase separation. Notably, our bottom-up approach reveals that a naive application of the Onsager currents yields an incorrect expression for the local dissipation.
\end{abstract}

\maketitle

%% ---- introduction ----

\section{Introduction}

\begin{figure*}
  \includegraphics[scale=1]{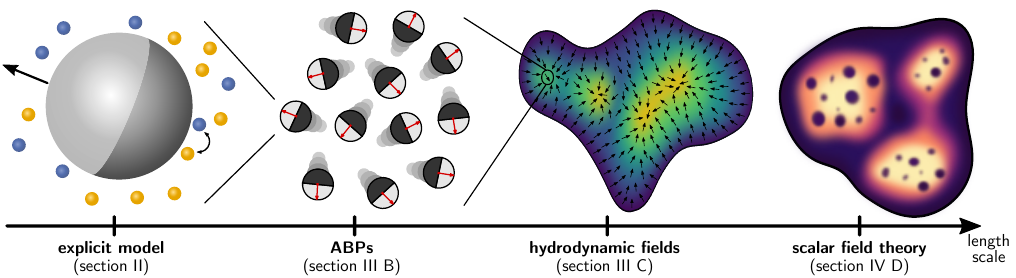}
  \caption{Schematic depiction of the studied length scales. Starting in Section~\ref{sec:explicit}, we consider single catalytic particles with explicit fuel molecules. In Section~\ref{sec:ABPs}, we establish a connection between this explicit model and a thermodynamically consistent derivation leading to the model of active Brownian particles (ABPs). We then turn to a macroscopic description in form of effective hydrodynamic fields in Section~\ref{sec:hydro}. Lastly, we eliminate the polarization field in Section~\ref{sec:scalar}, leaving us with a scalar field theory for the density alone.}
  \label{fig:scales}
\end{figure*}

%% - general introduction -

The development of macroscopic thermodynamics has in no small part been driven by the industrial revolution and its need for engines that exploit the transformation of heat into usable work~\cite{muller07}. A crucial step had been to recognize \emph{entropy} as the central constraint limiting the amount of extractable work. With the rapidly increasing capacity to image and manipulate systems on the micro- and even nanoscale, the scope of thermodynamics and statistical mechanics has vastly expanded. \emph{Entropy production} has since emerged as an important benchmark to characterize systems that are steadily driven away from thermal equilibrium~\cite{schnakenberg76, gaspard04, seifert05, parrondo09, landi21}. In particular living matter in the form of bacteria and cells must convert chemical energy to organize themselves and their environments in space and time, and to provide essential functions such as sensing and signaling~\cite{berg77, rieke98, rutherford12}, replication~\cite{kornberg05, england13, oriola18}, and locomotion~\cite{oster87, mitchison96, schwarz13}. 

%% - state of the art -

Basic features of living matter, such as locomotion, can also be realized in synthetic systems by harnessing a wide range of physical mechanisms that are studied within the field of (motile) active matter~\cite{marchetti13,bechinger16,gompper20}. Experimental realizations now routinely exploit diffusiophoresis~\cite{howse07, buttinoni13}, thermophoresis~\cite{jiang10}, photoactivation~\cite{palacci13, vutukuri20}, Quincke rotation~\cite{bricard13, das19}, electrophoresis~\cite{yan16}, and acoustophoresis~\cite{wang12} to realize persistent yet autonomous motion of individual agents. Irrespective of the exact microscopic details, any directed motion is necessarily accompanied by local time-reversal symmetry breaking and causes the continual dissipation of either residual or chemical free energy. While (motile) active matter undoubtedly falls into the realm of non-equilibrium systems, the majority of theoretical approaches remain founded in local equilibrium assumptions such as the determination of the self-propulsion speed from approximate solutions of the Stokes equation~\cite{anderson89,golestanian05,golestanian07, sabass10, sabass12, khair13, michelin14, wurger15,degraaf15,popescu16,ibrahim17, moran17, lisicki18}.

On a coarser (mesoscopic) scale, minimal models capture the persistent motion of active particles through a self-propulsion term directed along an orientation that undergoes rotational diffusion---so-called active Brownian particles (ABPs)~\cite{bechinger16, gompper20}. Suspensions of interacting motile particles have undergone substantial exploration. Of particular interest is the spontaneous aggregation of dense (liquid) domains surrounded by a dilute (gaseous) background even in the absence of cohesive forces, which is known as motility-induced phase separation (MIPS)~\cite{tailleur08, fily12, buttinoni13, palacci13, redner13, speck14, stenhammar14, cates15, liu19}. The goal of quantifying  collective behavior led to macroscopic field theories. Their derivation typically follows one of two routes: top-down approaches infer the phenomenological equations of motion from conservation laws and symmetries~\cite{toner98, marchetti13, wittkowski14, tjhung18}, whereas bottom-up approaches systematically coarse-grain the microscopic degrees of freedom into hydrodynamic fields and their evolution equations~\cite{dean96, bertin09, bialke13, speck20, speck22, pruessner22, vrugt23}. While top-down approaches enjoy great popularity due to their accessibility by virtue of an intuitive construction scheme, systematic bottom-up approaches have the advantage that they establish a connection between effective parameters of the field theory and parameters of the underlying microscopic model. Such a deeper understanding is crucial for the design of new material properties through tuning microscopic parameters.

Close to equilibrium in the (local) linear response regime, the framework of linear irreversible thermodynamics~\cite{groot84} can be used to derive equations of motion for coarse-grained degrees of freedom without explicit knowledge of the underlying microscopic dynamics. This approach dates back to the seminal work of Onsager~\cite{onsager31,onsager31a} and has a long history in the modeling of complex soft materials~\cite{doi11} such as liquid crystals under shear flow~\cite{olmsted90} and chemically driven biomolecular motors~\cite{julicher97}. Most importantly, transport properties due to small perturbations, such as thermal gradients and electrical fields, are encoded in equilibrium correlation functions. By postulating a linear coupling between the thermodynamic driving forces (commonly referred to as \emph{affinities}) and their conjugate macroscopic currents, irreversible thermodynamics provides a systematic route to the dissipation of near-equilibrium dynamics~\cite{groot84} and the derivation of thermodynamically consistent evolution equations.

Going beyond linear response, stochastic thermodynamics~\cite{sekimoto98, jarzynski11, seifert12, peliti21} has emerged as a comprehensive framework to treat systems dominated by fluctuations. In its essence, it methodically extends (typically macroscopic) thermodynamic notions to individual fluctuating trajectories of small (e.g., molecular-sized~\cite{seifert11, schmiedl07, speck21}) systems. Stochastic thermodynamics provides, as one of its hallmarks, fundamental connections between equilibrium and non-equilibrium quantities through the celebrated fluctuation relations~\cite{jarzynski97, jarzynski97a, crooks99, crooks00, hummer01}. A second hallmark is the equality of dissipated heat, as inferred from the stochastic first law, with the behavior of path probabilities under time reversal entering the second law of thermodynamics. More recent endeavors have focused on expanding the toolset of thermodynamic inference~\cite{seifert19}, where consistency constraints are enforced to derive experimentally accessible estimators of the entropy production which is notoriously hard to measure. Prominent examples include the thermodynamic uncertainty relation~\cite{barato15, horowitz17, horowitz20}, its numerous generalizations~\cite{gingrich17, koyuk19, dechant19, koyuk20, liu20, dieball23}, and novel approaches involving the waiting time statistics of partially observable (semi-)Markov models~\cite{skinner21a, vandermeer22, harunari22, vandermeer23}. Additionally, connections to macroscopic theories, where fluctuations play a subordinate role~\cite{leonard13, forastiere23, falasco23}, and systems with spatially extended correlations~\cite{falasco18, pruessner22, suchanek23, suchanek23a, venturelli23} are currently under scrutiny.

The framework of stochastic thermodynamics has found application in various attempts at quantifying the non-equilibrium character of active matter~\cite{speck16, fodor16, prawardadhichi18, gnesotto18, martinez19, szamel19, obyrne22}. To this end, a quantity that attracted immense interest, both from a theoretical~\cite{mandal17, nardini17, caprini18, shankar18, dabelow19, borthne20, fodor22, bowick22, agranov22, dabelow23} and experimental~\cite{battle16, seara18, ro22} perspective, is the irreversibility encoded in the time evolution of \emph{observable} degrees of freedom, also referred to as ``informatic'' or path entropy production~\cite{dabelow19, fodor22, agranov22, cates22}.

Only when including \emph{all} degrees of freedom that contribute to dissipation does the path entropy production equal the  \emph{thermodynamic} entropy production, which in turn relates to dissipation. Indeed, the standard recipes of stochastic thermodynamics applied to minimal active matter models~\cite{fodor16, mandal17, nardini17, shankar18, dabelow19, szamel19, nemoto19, fodor20, borthne20, caballero20, fodor22, rassolov22, ferretti22, davis23} lead to inconsistencies that have greatly obscured the correct identification of the dissipated heat. While an accurate thermodynamic theory for active matter is necessary to study energy balances in living matter~\cite{deng21,arunachalam23} and to design efficient active (heat) engines~\cite{krishnamurthy16, zakine17, pietzonka19, holubec20, kumari20, ekeh20, fodor21, malgaretti21, speck22a, datta22}, so far few attempts have been made at formulating thermodynamically consistent models of active agents~\cite{gaspard17, pietzonka18, speck18, gaspard18, gaspard19, fritz23, padmanabha23}.

The need for a thermodynamically consistent foundation becomes even more apparent at the level of field theories, which are, by construction, ignorant of microscopic details required for determining dissipation. At least close to equilibrium, it is thus tempting to conjecture the validity of linear irreversible thermodynamics in the presence of local drives that are characteristic for active matter~\cite{markovich21}. In theory, treating the locally supplied free energy as a small perturbation away from Boltzmann equilibrium paves the way to a linear thermodynamics, yielding expressions for the dissipation of active matter on different scales. While it has proven successful in the study of phoretically propelled Janus particles~\cite{gaspard20}, shown to reproduce the behavior of the cell cytoskeleton for hydrodynamic theories of active gels~\cite{kruse04,prost15,julicher18}, and applied to a broad class of active field theories~\cite{markovich21}, its universal applicability, nonetheless, remains an open question.

%% - our results -

Although coarse-graining necessarily sacrifices some information, carefully tracking the \emph{thermodynamically relevant} degrees of freedom during the coarse-graining procedure may allow a more rigorous underpinning of the thermodynamics of active field theories. Here we present and implement a bottom-up approach (cf. Fig.~\ref{fig:scales}) that starts from an explicit model involving molecular solutes (Sec.~\ref{sec:explicit}). The interconversion of these solutes at the surface of colloidal particles is shown to lead to self-propulsion, for which solute flux and force on the colloidal particle can be explicitly calculated. We then introduce an effective model, in which tight coupling between fuel consumption and directed displacement reproduces the explicit model. In Sec.~\ref{sec:tracking}, we track its dissipation through several coarse-graining steps: First, integrating out the solute molecules and expanding in a small parameter yields the evolution equation of interacting active Brownian particles together with exact expressions for the dissipation caused by each active particle. The calculated dissipation rate shows that active particle propulsion is accompanied by a constant solute flux between substrate and product reservoirs providing the work necessary to maintain directed motion. Second, going from particle-based to hydrodynamic fields yields as our central result an \emph{exact} expression for the dissipated heat, both local and averaged, in terms of the density and polarization fields.

Building on this result, in Sec.~\ref{sec:discussion}, we discuss ramifications and applications. Firstly, we explore the linear response regime and show how our results can be cast into a form that is consistent with the phenomenological framework of linear irreversible thermodynamics (Sec.~\ref{sec:LIT}). Secondly, we apply our result to illustrative examples in order to unveil the non-trivial thermodynamic footprint of confinement (Sec.~\ref{sec:wall}) and reveal the correct dissipation field of an active particle suspension undergoing MIPS (Sec.~\ref{sec:MIPS}). Lastly, we sketch a possible strategy to extend our results to scalar field theories (Sec.~\ref{sec:scalar}).

%% ---- explicit ----

\section{The explicit model: A single particle with fuel molecules}
\label{sec:explicit}

\subsection{Definition of model}

\begin{figure}[t]
    \centering
    \includegraphics{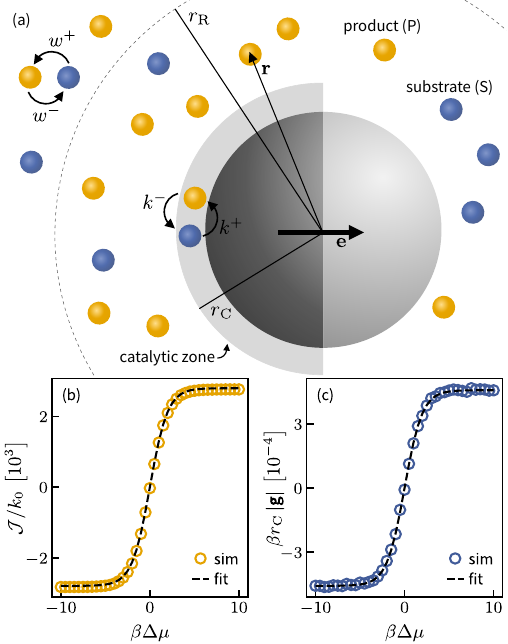}
    \caption{(a) Sketch of a colloidal Janus particle moving in the presence of two molecular solutes: substrate (blue) and product (yellow). One of the hemispheres is chemically inert and the other active (shaded zone), within which substrate can be converted into product and vice versa. In simulations, we implement a chemostat through a reservoir region for $r>r_\text{R}$, wherein interconversion occurs with rates $\om^\pm$. (b,\,c) Results of Brownian dynamics simulations involving a single stationary catalyst centered around the origin and $32\times10^5$ ideal reactants. Shown are (b) the chemical flux $\mathcal{J}$ and (c) the force magnitude $|\vec g|$ on a single catalytic particle as function of the chemical potential difference, verifying Eqs.~\eqref{eq:J-micro} and \eqref{eq:force_res}. Fitting our theoretical results, we obtain $\keff \approx 1396\, k_0$ from panel (b) and $\lam \approx 1.6\times 10^{-7}\, r_\text{C}$ from panel (c).}
    \label{fig:sketch}
\end{figure}

Our starting point is a single spherical colloidal particle submerged in an isothermal solvent at temperature $T$. For convenience, we define the inverse thermal energy $\beta\equiv(\kT)^{-1}$. The solvent is treated as an effective medium that gives rise to stochastic forces, the strength of which $D_0=\mu_0\kT$ is determined by the particle mobility $\mu_0$. The colloidal particle interacts with uncharged and much smaller (molecular) solutes, which we distinguish as either substrate (S) or product (P). For simplicity, we assume that both solute species have the same diffusion coefficient $D$ but interact through different isotropic potentials $u_\al(r)$ with the colloidal particle, where $\al$ is either S or P. Both potentials have a finite range $r<\rc$ beyond which the solutes diffuse freely. The surface of the colloidal particle is divided into inert and chemically active regions. Only within these active regions is the conversion between substrate and product possible due to a catalytic lowering of the activation barrier. Here we consider geometries with rotational symmetry defining the orientation $\vec e$, the simplest example for which is a Janus particle with two distinct hemispheres as depicted in Fig.~\ref{fig:sketch}(a).

In the following, we choose the center of the colloidal particle as the origin so that $\x$ denotes the position of a solute, $r=|\x|$ is its distance from the origin, and $\theta$ is the polar angle with the orientation $\vec e$. The conversion between substrate and product is a stochastic process described by two rates, $k^+$ for $\text{S}\to\text{P}$ and $k^-$ for $\text{P}\to\text{S}$. Both rates are zero except within the catalytic zone, where they obey the detailed balance condition
\begin{equation}
	\frac{k^+(r,\theta)}{k^-(r,\theta)} = e^{-\beta[u_\text{P}(r)-u_\text{S}(r)]} \equiv e^{-\beta\Delta u(r)},
	\label{eq:db}
\end{equation}
guaranteeing that the composite system will eventually reach thermal equilibrium. The equilibrium state has equal numbers of substrate and product solutes, as we have neglected any energy states internal to the reactants. Coupling to external reservoirs introduces a solute flux that drives the colloidal particle due to a non-uniform distribution of solute molecules.

Our model differs from the prevalent approach to model motility of catalytic Janus particles~\cite{golestanian05,golestanian07,cordova-figueroa08, khair13, wurger15, lisicki18}, which is through the hydrodynamic boundary layer approximation: the colloid is imagined to exert a force on nearby solute molecules, which propagates as a force density to the surrounding solvent~\cite{anderson89}. In this conventional approach, non-uniform current boundary conditions are then enforced at the particle surface leading to inhomogeneous concentration profiles (and thus gradients) that determine the self-propulsion speed. Such an independent treatment of surface potential and boundary conditions, however, violates local detailed balance and solvent flow cannot be driven by local concentration gradients alone~\cite{speck19}. Here, we present an alternative model without explicit solvent flows whereby particle-solute interactions are described via explicit pair potentials. This allows for an analytical solution---without introducing a slip velocity---and guarantees the correct stochastic energetics of the chemically driven system.

\subsection{Directed motion}
\label{sec:directed}

We now show that this model of explicit solute molecules exhibits directed motion. To this end, we consider an ideal chemostat that maintains constant densities $\bar c_\al$ far away from the colloidal particle ($r\to\infty$) with chemical potential difference $\Delta \mu\equiv\mu_\text{S}-\mu_\text{P}=\kT\ln(\bar c_\text{S}/\bar c_\text{P})$. The chemostat induces a flux of solutes from substrate to product reservoir, which performs work on the colloidal particle. The resulting speed is determined by the force due to the imbalance of solute concentrations on the different regions.

In Appendix~\ref{sec:thin}, we calculate the solute densities around the colloidal particle within a thin interaction layer approximation, i.e., solutes interact with the colloidal particle only within a spherical shell of effective thickness $\ell$ that is much smaller than the particle diameter. From these solute densities we extract the functional form of the total force
\begin{equation}
  \vec g = \frac{\ell k_0}{\beta D} h_1\bar c\mathcal{V}\tanh\left(\frac{\beta\Delta\mu}{2}\right)\hat{\vec e}_z
  \label{eq:force_res}
\end{equation}
acting on the colloidal particle with the total solute concentration $\bar c = \bar c_\text{S} + \bar c_\text{P}$. Here, $\mathcal V$ is akin to an effective volume [defined in Eq.~\eqref{eq:V}] independent of the chemostat and $h_1$ is a geometric coefficient quantifying the asymmetry between active and inert regions [defined in Eq.~\eqref{eq:h_l}]. Furthermore, we obtain the average net flux
\begin{equation}
  \mathcal J = 2k_\text{eff} \tanh{\left( \frac{\beta\Delta\mu}{2} \right)}
  \label{eq:J-micro}
\end{equation}
of solute particles from the substrate to product reservoir. Both of these quantities vanish in equilibrium where $\Delta\mu=0$ and $\bar c_\text{S}=\bar c_\text{P}=\bar c_\text{eq}$. The flux depends on the effective attempt rate
\begin{equation}
  k_\text{eff} \equiv 2 \pi \bar c \IInt{\bm\xi}{|\bm\xi|<\rc}{} \xi^2 k^-(\xi,\theta)e^{-\beta u_\text{P}(\xi)}.
  \label{eq:keff}
\end{equation}
The average propulsion speed $v_0=\mu_0|\vec g|=\lam\mathcal J$ along the orientation associated with the chemically induced displacements can thus be written as the flux (number of events per time) times an effective jump length
\begin{equation}
  \lam \equiv \frac{1}{2}\frac{D_0}{D}\frac{k_0}{k_\text{eff}}h_1\bar c\mathcal{V}\ell.
  \label{eq:lam}
\end{equation}
Each conversion thereby ``pushes'' the particle an effective distance $\lam$ along its orientation.

\subsection{Simulations}

We verify Eq.~\eqref{eq:force_res} for the force and Eq.~\eqref{eq:J-micro} for the flux by performing Brownian dynamics simulations in a finite volume $V$ with periodic boundaries. The box contains $32\times10^5$ non-interacting solutes. A single stationary colloid at the center of the box is represented by the external potentials $u_\text{S}(r)$ and $u_\text{P}(r)$. These potentials are short-ranged and repulsive, with interactions only occurring for distances $r < r_\text{C}$ (for details see Appendix~\ref{app:sim_explicit}).

We introduce a reservoir region $r>r_\text{R}$ [see Fig.~\ref{fig:sketch}(a)], wherein conversions take place with constant rates
\begin{equation}
  \frac{w^+}{w^-} = K e^{\beta \Delta\mu} = K \frac{\bar{c}_\text{S}}{\bar{c}_\text{P}}
  \label{eq:explicit-chemostat}
\end{equation}
obeying \emph{local} detailed balance~\cite{katz83, seifert12, vandenbroeck15}. The constant $K$ appearing here corrects for the finite box volume, which shifts the equilibrium chemical potential due to the presence of the catalyst ($K \to 1$ as $V \to \infty$) and is given in Appendix~\ref{app:sim_explicit}.

Within the catalytic zone $r < r_\text{C}$, we allow for conversions with rates
\begin{equation}
  k^\pm(r,\theta) = k_0 \,  h(\theta) \left[1+e^{\pm\beta \Delta u(r)}\right]^{-1}
  \label{eq:glauber-rates}
\end{equation}
that obey the detailed balance condition Eq.~\eqref{eq:db}. Here, $k_0$ is an attempt rate and $h(\theta)$ is an indicator function that is unity on active and zero on inert regions. We perform simulations with chemical potentials $\beta\Delta\mu \in [-10, 10]$ and calculate the solute flux [Fig.~\ref{fig:sketch}(b)] together with the total force [Fig.~\ref{fig:sketch}(c)]. Fitting the analytical expression for the solute flux [Eq.~\eqref{eq:J-micro}] for $\keff$ we find, in excellent agreement with simulation data, $\keff \approx 1396\,k_0$ as the effective attempt rate. Plugging this into the expression for the force [Eq.~\eqref{eq:force_res}], we further obtain $\lam \approx 1.6\times 10^{-7}\, r_\text{C}$ by fitting the simulated forces. With $r_\text{C}$ comparable to the radius of a colloidal particle (i.e., hundreds of \si{nm} to a few \si{\micro m}), the obtained displacement length is of the order $\sim 0.01$ to $0.1\,\text{pm}$ in reasonable agreement with previous studies based on boundary layer approaches for phoretically propelled particles~\cite{golestanian05, sabass12, ebbens12}. Moreover, since $\keff$ implicitly depends of the interaction layer thickness $\ell$ due to the limited support of the integral in Eq.~\eqref{eq:keff}, we recover the commonly reported $v_0 \propto \lam\keff \sim \ell^2$ scaling of the self-propulsion speed~\cite{golestanian05, sabass12, ebbens12, speck19}.

%% ---- tracking ----

\section{Tracking dissipation}
\label{sec:tracking}

\subsection{Effective chemical events}
\label{sec:eff_chem}

\begin{figure}
  \includegraphics[scale=1]{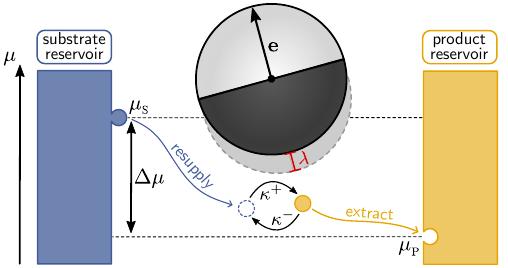}
  \caption{Schematic illustration of a free catalytic particle undergoing effective chemical events. After each chemical event, converting substrate into product, the particle is displaced by a small distance $\lam$ along its orientation $\vec e$. To assure ideal substrate and product concentrations external chemostats instantly resupply/remove the used and produced reaction constituents from/to (infinite) reservoirs. If the backward reaction takes place instead, this routine is reversed and the particle is displaced against its orientation.}
  \label{fig:chemiostat}
\end{figure}

The central lesson of the previous section (Sec.~\ref{sec:explicit}) has been that we can map the explicit model onto effective chemical events with attempt rate $k_\text{eff}$. While such an effective model has been proposed earlier~\cite{speck18}, we have shown here that it follows naturally from a more microscopic description involving the conversion of fuel molecules. In the effective model, we assume that chemical reactions between solute and product molecules are catalyzed on the colloidal surface. Free energy $\Delta \mu$ is either liberated ($\text{S} \to \text{P}$) or consumed ($\text{P} \to \text{S}$) depending on the direction. Moreover, each reaction coincides with a displacement of small but finite length $\lam$, which is known as \emph{tight coupling}. While we have obtained an explicit expression [Eq.~\eqref{eq:lam}], in the following we will treat $\lam$ as a parameter of the effective model. Clearly, the chemical events are stochastic in nature, allowing us to model their dynamics as an effective two-state model where transitions occur with rates $\kap^+$ in the forward and $\kap^-$ in the backward direction (see Fig.~\ref{fig:chemiostat}). The detailed balance condition [Eq.~\eqref{eq:db}] is then replaced by
\begin{equation}
  \frac{\kap^+}{\kap^-} = e^{\beta\Delta\mu},
  \label{eq:LDB_free}
\end{equation}
assuming that the reservoirs of solute and product molecules are ideal in the sense that concentrations, and thus the driving affinity $\Delta\mu$, remain constant at all times.

The evolution equation for the probability $\psi(\x,t)$ of the particle position $\x$ reads
\begin{equation}
  \pd{\psi}{t} = -\nabla\cdot(v_0\vec e\psi) + D^\text{c}\nabla^2\psi
\end{equation}
in the limit $\lam\to 0$ of vanishing displacement. Here
\begin{equation}
  v_0 \equiv \lam (\kap^+ - \kap^-)= 2k_\text{eff}\lam\tanh{\left(\frac{\beta\Delta\mu}{2}\right)}
  \label{eq:v0}
\end{equation}
is the self-propulsion speed, where in the second step we assume symmetric rates
\begin{equation}
  \kap^+ = 2 k_\text{eff} \frac{\bar c_\text{S}}{\bar c}, \qquad
  \kap^- = 2 k_\text{eff} \frac{\bar c_\text{P}}{\bar c}
  \label{eq:kap}
\end{equation}
obeying Eq.~\eqref{eq:LDB_free}. This speed agrees exactly with the result obtained for the explicit model. At quadratic order in $\lam$, the chemical events alone induce translational diffusion with coefficient
\begin{equation}
  D^\text{c} \equiv \frac{\lam^2}{2}(\kap^+ + \kap^-)  = k_\text{eff}\lam^2,
  \label{eq:Dc}
\end{equation}
which has no dependence on $\Delta \mu$ and scales as $D^\text{c} \propto \bar{c}$ [cf. Eq.~\eqref{eq:keff}].

%%% ---- ABPs ----

\subsection{The mesoscale: Active Brownian particles}
\label{sec:ABPs}

\subsubsection{A thermodynamically consistent many-body model}
\label{sec:many}

We now extend the model of effective chemical events to $N$ interacting active particles. The full state of the system is described by the position $\x_i$ and orientation $\vec e_i$ of each particle $i\in \lbrace1,\ldots, N\rbrace$ together with the net number $n_i$ of chemical events. The $N$-body joint distribution $\Psi\left(\vec{r}^N, \vec{e}^N, n^N; t\right)$ then obeys the exact evolution equation
\begin{equation}
  \pd{\Psi}{t} = -\sum_{i=1}^N \left\{\nabla_i \cdot \vec j^\text{p}_i - D_\text{r}\Delta_{\vec e_i}\Psi + \widehat\nabla_i j^\text{c}_i \right\}.
  \label{eq:nbody}
\end{equation}
Each active particle is subjected to three stochastic processes. First, positional changes due to translational diffusion subject to interparticle and external forces give rise to the particle currents
\begin{equation}
  \vec j^\text{p}_i \equiv \mu_0(\vec F^\text{ex}-\nabla_i U)\Psi - D_0\nabla_i\Psi.
  \label{eq:cur_p}
\end{equation}
The potential energy $U(\vec r^N)\equiv\sum_{i<j}u(|\vec r_i - \vec r_j|)$ is composed of pair interactions with (colloidal) pair potential $u(r)$ and $\Fex$ is a position-independent, externally applied force. Extension to position-dependent, conservative external forces is straightforward. Second, particle orientations undergo free rotational diffusion with diffusion coefficient $D_\text{r}$ and $\Delta_{\vec e_i}$ denotes the spherical Laplace operator in $d$ dimensions that acts on the orientation of the ith particle. Third, particles are displaced by the distance $\lam$ along their orientation during each chemical reaction (Fig.~\ref{fig:chemiostat}). These active translations are captured by the discrete difference operator
\begin{equation}
  \widehat\nabla_i j^\text{c}_i \equiv j^\text{c}_i(\vec r_i,n_i) - j^\text{c}_i(\vec r_i-\lam\vec e_i,n_i-1),
  \label{eq:diff_op}
\end{equation}
where we have defined the particle-resolved chemical currents
\begin{equation}
  j^\text{c}_i \equiv \Psi(\vec{r}_i, n_i) k_i^+(\vec{r}_i) - \Psi(\vec{r}_i + \lam \vec{e}_i, n_i+1)k_i^-(\vec{r}_i+\lam\vec e_i).
  \label{eq:ch_cur}
\end{equation}
To improve readability, only the arguments that change in transitions are indicated. These chemical currents simultaneously convey changes in particle positions (along their orientations) and the net number of chemical events. As will be seen later, they give rise to the chemical flux $\mathcal J$ between substrate and product reservoirs, which constitutes the work input.

\subsubsection{Dissipated heat}

We can now apply the standard framework of stochastic thermodynamics to identify the dissipation. To this end, it is instructive to consider infinitesimal increments for a single active particle. The first law of thermodynamics reads $\dbar q=\dbar w-\dd U$, where throughout we adopt the convention that work applied to the system and heat dissipated into the surrounding medium take positive values. We split the stochastic first law for each particle into (i)~displacements caused by the total force $\Fex-\nabla_i U$ and (ii)~chemically induced displacements along its orientation. For (i), work $\dbar w^\text{p}_i=\Fex \circ \dd\vec r_i^\text{p}$ is performed by the external force with infinitesimal displacement $\dd \vec r_i^\text{p}$. Throughout we work in the Stratonovich convention~\cite{stratonovich66} and denote corresponding (inner) products by ``$\circ$''. The associated change in internal energy $U$ is captured by the shift in potential energy $\dd U^\text{p}_i=\nabla_i U\circ \dd\vec r_i^\text{p}$, from which follows the dissipated heat
\begin{align}
  \dbar q^\text{p}_i = \dbar w_i^\text{p}-\dd U_i^\text{p}
  = (\Fex -\nabla_i U) \circ \dd \vec r_i^\text{p}.
\end{align}
The extension to $N$ particles is straightforward and, following the standard procedure of stochastic thermodynamics~\cite{seifert12}, allows us to express the average dissipated heat per unit of time
\begin{equation}
  \dot Q^\text{p} \equiv \sum_{i=1}^N \Int{\vec{r}^N \dd \vec{e}^N \dd n^N } \vec{j}^\text{p}_i \cdot (\vec F^\text{ex} - \nabla_i U)
  \label{eq:pepr_general}
\end{equation}
stemming from particle displacements due to the forces exerted onto them. To prevent notational clutter, $\Int{\vec r^N \dd \vec e^N \dd n^N}$ denotes integration over positions, orientations, and summation over the number of chemical events across all $N$ particles.

For the chemical events, the work increment splits into two contributions, $\dbar w^\text{c}_i = \Delta\mu + \lam \vec e_i \cdot \Fex$. The first contribution describes the work injected in the form of chemically liberated free energy, whereas the second captures the mechanical work to displace the particle by a short distance $\lambda$ along its orientation. The corresponding change in internal energy reads $\dd U^\text{c}_i = U(\vec r_i +\lam \vec e_i) - U(\vec r_i)$, which results in the dissipated heat associated with a single chemical event
\begin{equation}
  \dbar q^\text{c}_i = \Delta \mu + \lam \vec e_i \cdot \Fex - U(\vec r_i +\lam \vec e_i) + U(\vec r_i).
\end{equation}
This dissipation constrains the forward and backward rates through the local detailed balance condition [cf. Eq.~\eqref{eq:LDB_free}]
\begin{equation}
  \frac{k_i^+(\vec r_i)}{k_i^-(\vec r_i + \lam \vec e_i)} = e^{\beta \dbar q^\text{c}_i},
  \label{eq:ldb}
\end{equation}
one of the cornerstones of stochastic thermodynamics.
In analogy with Eq.~\eqref{eq:pepr_general}, we thus identify 
\begin{equation}
  \dot Q^\text{c} \equiv \sum_{i=1}^N \Int{\vec{r}^N \dd  \vec{e}^N \dd n^N} j^\text{c}_i \kT\ln\frac{k_i^+}{k_i^-}
  \label{eq:cepr_general}
\end{equation}
as the average heat rate caused by (local) dissipation of chemically liberated free energy and the associated particle translations along their respective orientations. The total dissipation rate of the system is given by the superposition $\dot Q=\dot Q^\text{p}+\dot Q^\text{c}$. Our result holds for idealized chemostats, i.e., constant $\Delta \mu$. In case of non-constant affinity, chemical reservoirs undergo a change in entropy yielding an additional contribution to the dissipation (for a comprehensive discussion see Ref.~\cite{seifert11}). Notably, rotational diffusion does not contribute to the dissipation due to its time-reversal symmetry. 

\subsubsection{Coarse-graining chemical events}

Inspecting the results of the previous section (Sec.~\ref{sec:many}) closer, we see that the net number of chemical events $n_i$ are needed for bookkeeping but can be eliminated due to the tight coupling between such events and active displacements. To this end, we expand
\begin{equation}
  \Psi(\vec{r}_i \pm \lam \vec{e}_i, n_i \pm 1) \approx \Psi(\vec{r}_i \pm \lam \vec{e}_i, n_i) \pm \partial_{n_i}\Psi(\vec{r}_i \pm \lam \vec{e}_i, n_i)
\end{equation}
assuming real valued $n_i$ for simplicity. Note that in doing so, we implicitly invoke two assumptions:~(i) The time increment of consecutive observations $\dd t$ is required to be sufficiently small, such that reaction coordinates $\{n^N\}$ are essentially statistically independent Poisson random variables. (ii)~Simultaneously, $\dd t$ has to be long enough to assure the occurrence of a large number of reactions within $[t, t+\dd t]$. Upon satisfying the latter condition, integer Poisson variables are well-approximated by real-valued Gaussian random variables, by virtue of the central limit theorem~\cite{gillespie00}. Upon integrating
\begin{equation}
  \psi\left(\vec{r}^N, \vec{e}^N; t \right) = \Int{n^N} \Psi\left(\vec{r}^N, \vec{e}^N, n^N; t\right),
  \label{eq:psi}
\end{equation}
the state of the system is described by the collection $\lbrace\vec{r}^N, \vec{e}^N \rbrace$ of particle positions and orientations with joint probability $\psi\left(\vec{r}^N, \vec{e}^N; t\right)$.

To simplify the evolution equation for $\psi$ further, we now exploit that $\lam$ is a microscopic length scale much smaller than the size of the colloidal particles (cf.~Sec.~\ref{sec:explicit}). To proceed, we require explicit expressions for the transitions rates, for which we again choose a symmetric form
\begin{equation}
  k_i^\pm \equiv \kap^\pm e^{\frac{\beta}{2}\left[- U(\vec{r}_i \pm \lam \vec{e}_i) + U(\vec{r}_i) + \lam\vec e_i\cdot\vec F^\text{ex}\right]},
  \label{eq:rates}
\end{equation}
with rates $\kap^\pm$ for a free particle [Eq.~\eqref{eq:kap}] so that local detailed balance [Eq.~\eqref{eq:ldb}] is satisfied. We note that an alternative choice of rates, similar to the one in Eq.~\eqref{eq:glauber-rates}, leaves the results of the following calculation unchanged (for details see Appendix~\ref{app:rates}).

We expand the joint distribution [Eq.~\eqref{eq:psi}] and transition rates [Eq.~\eqref{eq:rates}] in orders of $\lam$ to obtain
\begin{multline}
  \psi(\vec{r}_i \pm \lam \vec{e}_i, \vec{e}_i) = \psi(\vec{r}_i, \vec{e}_i) \pm \lam \vec{e}_i \cdot \nabla_i \psi(\vec{r}_i, \vec{e}_i) \\ + \frac{\lam^2}{2} (\vec e_i \cdot \nabla_i)^2 \psi(\vec r_i) + \mathcal{O}(\lam^3)
  \label{eq:expans_psi} 
\end{multline}
and
\begin{align}
  k_i^\pm(\vec{r}_i \mp \lam \vec{e}_i) &= k_i^\pm(r_i) \mp \lam\vec e_i \cdot \nabla_i k_i^\pm + \mathcal{O}(\lam^3).
  \label{eq:expans_k}
\end{align}
Plugging these expansions into the definition of the discrete difference operator [Eq.~\eqref{eq:diff_op}], we find
\begin{align}
  \widehat\nabla_i j^\text{c}_i &=  \lam \vec{e}_i \cdot \nabla_i \left[ (k_i^+ - k_i^-)  \psi - \frac{1}{2} \lam (k_i^+ + k_i^-)\vec e_i \cdot \nabla_i \psi \right] %+ \mathcal{O}(\lam^3)
  \nonumber \\
  &= \lam \vec{e}_i \cdot \nabla_i j_i^\text{c}
  \label{eq:discrete_op}
\end{align}
to order $\propto \lam^2$, where the second line defines the chemical current
\begin{equation}
  j_i^\text{c} \equiv \frac{v_i}{\lam} \psi - \frac{D^\text{c}}{\lam} \vec{e}_i \cdot \nabla_i \psi + \mathcal{O}(\lam^2)
  \label{eq:ccur_ABPs}
\end{equation}
in the limit of small displacement length $\lam$. Note that the speed $v_i\equiv v_0+\mu^\text{c}\vec e_i\cdot(\vec F^\text{ex}-\nabla_i U)$, with constant speed $v_0$ [Eq.~\eqref{eq:v0}] and chemical mobility $\mu^\text{c}\equiv\beta D^\text{c} \propto \lam^2$ [Eq.~\eqref{eq:Dc}], acquires a non-constant, position-dependent contribution. To arrive at Eq.~\eqref{eq:ccur_ABPs}, we further expand the transition rates, given in Eq.~\eqref{eq:rates}, as
\begin{equation}
  k_i^\pm(\vec{r}_i) = \kap^\pm \left[1 \pm \lam\frac{\beta}{2} \vec{e}_i \cdot ( \vec F^\text{ex} - \nabla_i U) \right] + \mathcal{O}(\lam^2)
\end{equation}
to leading order in $\lam$ and plug them into Eq.~\eqref{eq:expans_k}.

By keeping terms to linear order of $\lam$ only, Eq.~\eqref{eq:nbody} reduces to the Smoluchowski equation
\begin{multline}
  \pd{\psi}{t} = -\sum_{i=1}^N \left\{ \nabla_i \cdot \left[ v_0\vec{e}_i +\mu_0(\vec F^\text{ex} -\nabla_i U)-D_0\nabla_i  \right] \right. \psi \\ \left. - D_\text{r} \Delta_{\vec e_i}\psi \right\}.
  \label{eq:abp}
\end{multline}
This is the evolution equation of active particles that are propelled with constant speed $v_0$. In the literature, this model is commonly referred to as active Brownian particles (ABPs), a well-studied paradigm in active matter research~\cite{romanczuk12, fily12, buttinoni13, palacci13, cates15, bechinger16, siebert18, digregorio18, gompper20}. The novelty of our approach is that information about the underlying self-propulsion mechanism directly enters through the self-propulsion speed $v_0$ in the form of the effective attempt rate $k_\text{eff}$, the chemical potential difference $\Delta \mu$, and the displacement length $\lam$. Moreover, necessary thermodynamic details are directly encoded in the chemical currents [Eq.~\eqref{eq:ccur_ABPs}], which we exploit in the following to extract the exact dissipation rate.

\subsubsection{Mesoscopic dissipation rate}

We now consider steady states defined by a time-independent joint probability $\psi(\x^N,\vec e^N)$. To calculate the dissipation rate, we insert the particle current [Eq.~\eqref{eq:cur_p}] and the chemical current [Eq.~\eqref{eq:ccur_ABPs}] into the respective contribution to the total dissipation [Eqs.~\eqref{eq:pepr_general} and \eqref{eq:cepr_general}]. The first takes the form
\begin{equation}
  \dot Q^\text{p} = \sum_{i=1}^N \Int{\vec{r}^N \dd \vec{e}^N}  \left[\mu_0 (\vec F^\text{ex} - \nabla_i U)^2\psi - D_0 (\nabla_i^2 U) \psi\right],
  \label{eq:pepr_abp}
\end{equation}
which is generally non-zero except in thermal equilibrium when $\psi$ is Boltzmann-distributed. The associated rate of dissipation due to effective chemical events 
\begin{multline}
  \dot Q^\text{c} = \sum_{i=1}^N \Int{\vec{r}^N \dd \vec{e}^N} \left[ \frac{v_0}{\lam}\Delta \mu \psi + v_0\vec e_i\cdot(\vec F^\text{ex} -\nabla_i U)\psi \right. \\
  \left. + \frac{\mu^\text{c}}{\lam}\Delta\mu \vec{e}_i \cdot (\vec F^\text{ex} - \nabla_i U)\psi 
  -\Delta \mu \frac{D^\text{c}}{\lam} \vec{e}_i\cdot\nabla_i\psi  \right]
  \label{eq:cepr_abp}
\end{multline}
follows by additionally plugging the local detailed balance condition [Eq.~\eqref{eq:ldb}] into Eq.~\eqref{eq:cepr_general} and discarding terms beyond linear order in $\lam$ to be consistent with the limit in which ABPs emerge. 

To write the total average dissipation rate $\dot Q$ in a more compact way, we eliminate terms that do not contribute to a global average. To this end, we follow Ref.~\cite{pietzonka18} and consider the Smoluchowski equation [Eq.~\eqref{eq:abp}] under the assumption of stationarity ($\partial_t \psi = 0$),
\begin{multline}
  0 = -\sum_{i=1}^N \{ \nabla_i \cdot \left[ v_0\vec{e}_i +\mu_0(\vec F^\text{ex} - \nabla_i U) -D_0\nabla_i  \right] \psi \\ -D_\text{r} \Delta_{\vec e_i} \psi \}.
  \label{eq:smol_stat}
\end{multline}
Multiplication of each sum element by $U-\vec r_i \cdot \vec F^\text{ex}$, followed by repeated integration by parts, results in the condition
\begin{multline}
  0 = -\sum_{i=1}^N \Int{\vec{r}^N \dd \vec{e}^N} \left[ v_0 \vec e_i\cdot(\vec F^\text{ex} -\nabla_i U)\psi \right. \\ \left.+ \mu_0(\vec F^\text{ex} - \nabla_i U)^2 \psi 
  -  D_0 (\nabla_i^2 U)\psi   \right],
  \label{eq:condition}
\end{multline}
where we dropped surface terms that vanish for suitably chosen boundary conditions (cf.\ Appendix~\ref{app:BC}). Rearranging terms and subsequent insertion in Eq.~\eqref{eq:cepr_abp} together with Eq.~\eqref{eq:pepr_abp} permits us to write the total rate of heat dissipation as 
\begin{equation}
  \dot Q = \Delta\mu \sum_{i=1}^N \Int{\vec{r}^N \dd \vec{e}^N} j_i^\text{c} \Delta \mu
  \equiv \Delta\mu \sum_{i=1}^N \mathcal{J}_i.
  \label{eq:epr:ABPs}
\end{equation} 
Here we identified the solute flux due to the $i$th particle as $\mathcal{J}_i \equiv \int\dd \vec r^N \dd \vec e^N (v_i/\lam)\psi$ by dropping boundary terms stemming from the integration of Eq.~\eqref{eq:ccur_ABPs}, which vanish for hard walls and in infinite or periodic systems. Importantly, we thus find that the dissipated heat is given by the net number of chemical conversions $\mathcal J=\sum_i\mathcal J_i$ times the liberated free energy $\Delta\mu$ per event. While this result should have been anticipated, we have shown it explicitly from the definition of the dissipated heat and the evolution of the joint probability. We emphasize that both flux and speed are related by $\mean{v_i}=\lam\mathcal J_i$. Accordingly, the entropy production is solely determined by effective chemical events coupled to particle displacement, as all translational diffusion terms cancel.

%% ---- hydrodyn ----

\subsection{The hydrodynamic scale: Dissipation from fields}
\label{sec:hydro}

\subsubsection{Coarse-graining}
\label{sec:cg}
Thus far, we restricted ourselves to the discussion of particle-resolved models. Now, we want to go one step further by turning towards a field-theoretic description of both the dynamics and thermodynamics of the system at hand by deriving effective hydrodynamic equations while simultaneously extracting the corresponding thermodynamically consistent dissipation rate. To this end, we employ the explicit coarse-graining scheme introduced by Bialké et al.~\cite{bialke13}.

Starting with the evolution equation of the joint distribution in Eq.~\eqref{eq:abp}, we adopt an effective one-particle picture by integrating over positions and orientations of all but one (tagged) particle. Since all particles are assumed to be identical, we designate $i=1$ as the tagged particle and drop subscripts from here on out. Subsequently, the equation of motion of the marginal one-point particle density $\psi_1(\vec{r},\vec{e};t)$ takes the form
\begin{align}
  \pd{\psi_1}{t} &= N \Int{\vec{r}^{N-1} \dd \vec{e}^{N-1}} \pd{\psi}{t} \nonumber 
  \\
  &= -\nabla \cdot \left[ v_0 \vec{e} + \mu_0 (\vec{F}^\text{ex} + \vec{F}) - D_0\nabla \right] \psi_1 + D_\text{r} \Delta_{\vec e} \psi_1.
  \label{eq:smol_one_point}
\end{align} 
Here, $\vec{F}$ describes the forces exerted on the tagged particle due to particle interactions. In the case of pairwise interactions, these couple to $\psi_2(\vec r^\prime | \vec r, \vec e; t)$, the density to find a second particle at position $\vec r^\prime$ conditioned on the tagged particle being at $\vec r$ with orientation $\vec e$ at time $t$. Thus, the (conditional) interaction forces are given by
\begin{equation}
  \vec{F} = \Int{\vec r^\prime} \vec{f}(\vec r - \vec r^\prime) \psi_2(\vec r^\prime | \vec r, \vec e; t)
  \label{eq:force_cond}
\end{equation}
with two-body force $\vec f=-\nabla u$ for the pair potential $u(r)$ between the active colloidal particles. To complete the coarse-graining, we introduce order parameter fields as moments of the particle orientation, namely the local density $\rho(\vec r,t)\equiv\Int{\vec e}\psi_1$ and the polarization density $\vec p(\vec r,t)\equiv\Int{\vec e}\vec e\psi_1$, alongside higher moments; and calculate their evolution equations by plugging Eq.~\eqref{eq:smol_one_point} into the time derivatives of their respective definition. In doing so, we obtain an infinite hierarchy of coupled equations~\cite{saintillan15}, reminiscent of the BBGKY hierarchy in kinetic theory~\cite{hansen13}. The density obeys the continuity equation
\begin{equation}
  \pd{\rho}{t} = -\nabla \cdot \vec j
  \label{eq:eom_density}
\end{equation}
with one-body particle current
\begin{equation}
  \vec j(\x,t) = v_0 \vec{p} +  \mu_0 (\rho\vec F^\text{ex} + \mean{\vec{F}}_{\vec{r}}) - D_0\nabla \rho.
  \label{eq:cur_p_macro}
\end{equation}
Here, $\mean{\cdot}_{\vec r}$ denotes averages over the tagged particle orientation. In contrast, higher-order fields, like the local polarization density following the evolution equation
\begin{multline}
  \pd{\vec p}{t} = -\nabla \cdot \left[ \frac{v_0}{d}\rho\id + v_0 \vec Q +\mu_0(\vec p \vec F^\text{ex} + \mean{\vec e \vec F}_{\vec r}) -D_0 \nabla \vec p  \right] \\ - D_\text{r} \vec p,
  \label{eq:eom_pol}
\end{multline}  
are generally non-conserved and decay increasingly fast with characteristic decay rates $\propto D_\text{r}$~\cite{speck21}. Here, $\vec Q = \Int{\vec e} (\vec e \vec e - \id/d)$ is the nematic tensor and we denote direct products as $(\vec a \vec b)_{ij} = a_i b_j$. Since each evolution equation couples to the next higher-order moment, an approximate closure is needed to truncate the hierarchy of equations. For the dissipation, however, general results are attainable at the level of density and polarization without introducing further assumptions.

\subsubsection{Macroscopic dissipation rate}

To arrive at an expression for the steady-state dissipation rate of such macroscopic fields, we repeat the exact same coarse-graining procedure for the global dissipation rate of ABPs given by Eq.~\eqref{eq:epr:ABPs}. In contrast to the previously discussed particle-resolved dynamics, where extending the domain over the full system implied vanishing boundary terms (cf. Appendix~\ref{app:BC}), treatment on the level of hydrodynamic equations enables the study of subsystems without contradicting mass conservation. 

\begin{figure}[t]
  \includegraphics[scale=1]{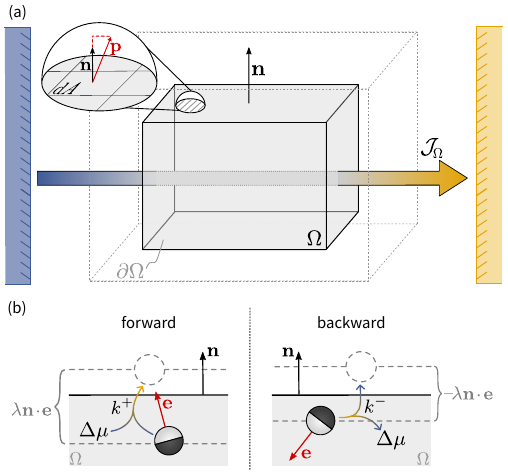}
  \caption{Subsystems and boundary fluxes. (a)~Sketch of a subsystem $\Omega$ bounded by the surface $\partial \Omega$. The shaded arrow indicates the average chemical flux $\mathcal{J}_\Omega$ from the substrate reservoir (blue, left) to the product reservoir (yellow, right). The red arrow indicates the polarization $\vec p$ on the boundary. Projecting $\vec p$ onto the outwards facing normal vector $\vec n$ and integrating over the infinitesimal surface element $\dd A$ gives the boundary contribution $\dot \Gamma_\Omega$ for the dissipation rate. (b)~Caricature illustrating the effect of a boundary on the dissipation rate. Every forward reaction (left) that pushes the particle out of $\Omega$ transfers $\Delta \mu$ to the surrounding medium. Backward reactions (right) replenish a $\Delta \mu$ to the chemostat and associated boundary crossings are not accompanied by heat transport. The difference of both processes (and the analog for particles entering the subdomain) leads to the average boundary flux Eq.~\eqref{eq:boundary}.}
  \label{fig:subsystem}
\end{figure}

The macroscopic dissipation rate associated with a (sub-)region $\Om$ [cf.~Fig.~\ref{fig:subsystem}(a)] reads 
\begin{equation}
  \dot Q_\Om = \IInt{\vec{r}}{\Om}{} j^\text{c}(\vec r) \Delta \mu = (\mathcal J_\Om - \dot \Gamma_\Om)\Delta \mu
  \label{eq:epr:field}
\end{equation}
with macroscopic local chemical current 
\begin{equation}
  j^\text{c}(\vec r) = \frac{v_0}{\lambda} \rho + \frac{\mu^\text{c}}{\lam} ( \vec p \cdot \vec F^\text{ex} + \langle \vec e \cdot \vec F \rangle_{\vec r}) - \frac{D^\text{c}}{\lam}  \nabla \cdot \vec p
  \label{eq:cur_c_macro}
\end{equation}
obtained from Eq.~\eqref{eq:ccur_ABPs}. The chemical flux
\begin{equation}
  \mathcal J_\Om = \IInt{\vec{r}}{\Om}{} \left[ \frac{v_0}{\lam} \rho + \frac{\mu^\text{c}}{\lam} (\vec p \cdot \vec F^\text{ex} + \mean{\vec{e}\cdot\vec{F}}_{\vec{r}}) \right]
  \label{eq:flux:field}
\end{equation}
contributing the bulk dissipation is the same as inferred from Eq.~\eqref{eq:epr:ABPs} (restricted to $\Om$). The second term
\begin{equation}
  \dot\Gamma_\Om = \frac{\lam}{2}(\kap^++\kap^-) \OInt{A}{\partial\Om}{} \vec n \cdot \vec p
  \label{eq:boundary}
\end{equation}
involves a generally non-zero integral (unless $\Om$ is the full system) along the subsystem boundary $\partial \Om$ and captures the mismatch between the work extracted from the chemostat and dissipation due to directed particle motion.

That a boundary term in the form of Eq.~\eqref{eq:boundary} arises can be rationalized as follows: Assume that particles located at the boundary $\partial\Om$ are aligned with the normal vector $\vec n$, i.e., $\dot \Gamma_\Omega > 0$. In order to perform a reactive step particles extract work from the chemostat while still inside $\Omega$. However, by performing the step they exit the subsystem, thereby transferring this energy into the surroundings as heat. As a result, the amount of heat dissipated by $\Omega$ decreases by the exact rate of effectively transferred heat $\dot Q^\text{tr}_\Omega = \dot \Gamma_\Omega \Delta \mu$, maintaining energy conservation of the composite system. Analogously, for $\dot \Gamma_\Omega < 0$ the dissipation rate of $\Omega$ increases. Note that we recover the first law of thermodynamics on the level of active fields since Eq.~\eqref{eq:epr:field} equates the rate of heat dissipation in any subsystem $\dot Q_\Omega$ with the difference between the rate of work $\dot{\mathcal{W}}^\text{in}_\Omega=\mathcal{J}_\Omega \Delta \mu$ injected by the chemostat and the rate $\dot Q^\text{tr}_\Omega = \dot \Gamma_\Omega \Delta \mu$ with which heat is transferred between $\Omega$ and its surrounding. Since the last contribution is solely determined by local polar order at the subsystem boundary [see Eq.~\eqref{eq:boundary}], its presence (or absence) generally relies on the system geometry and the specific choice of subsystem $\Omega$ (see Sec.~\ref{sec:wall}).

Extracting a \emph{local dissipation rate} $\dot q(\vec r)$, we first note that there is an apparent gauge freedom $\dot q\to\dot q+\nabla\cdot\vec A$ that does not change the total dissipation $\dot Q$ and, for finite subsystems $\Om$, modifies the boundary contribution $\dot\Gamma_\Om$. However, following our argument above the boundary term [Eq.~\eqref{eq:boundary}] exactly carries the dissipation due to particles crossing the boundary during a chemical event [see Fig.~\ref{fig:subsystem}(b)] and we conclude that $\vec A=0$. The local dissipation rate is thus
\begin{equation}
  \dot q(\vec r) = j^\text{c}(\vec r) \Delta \mu
  \label{eq:lepr:fields}
\end{equation}
with $j^\text{c}(\vec r)$ as given in Eq.~\eqref{eq:cur_c_macro}. As we illustrate in Sec.~\ref{sec:illustrations}, Eq.~\eqref{eq:lepr:fields} allows one to unveil inhomogeneities in the dissipation rate due to local features, such as, e.g., confinements or interfaces between coexisting phases.

%% ---- discussion ----

\section{Discussion}
\label{sec:discussion}

\subsection{Tight coupling}

Our central result is the systematic identification of the solute flux $\mathcal J$ [through Eqs.~\eqref{eq:epr:ABPs} and \eqref{eq:flux:field}] between the two reservoirs that is mediated by the active particles. This flux is tightly coupled to the particle speeds due to the fact that each chemical event is necessarily accompanied by an active translation of the particle. One might thus wonder how much of this flux could have been inferred from the evolution of particles alone, e.g., from the evolution equation~\eqref{eq:abp} that only depends on a constant speed $v_0$ and is agnostic to the underlying mechanism that generates the propulsion speed. Within stochastic thermodynamics there are essentially two interpretations for the term $v_0\vec e_i$: as an effective force or as an external flow field. These are distinguished by their behavior under time reversal (even vs.~odd). Unsurprisingly, neither interpretation yields the correct result.

For small $\lam$ after expanding to linear order, both the speeds and the solute flux [Eq.~\eqref{eq:flux:field}] can be written as two contributions. The first contribution only depends on the constant speed $v_0$ while the second contribution involves particle interactions through the correlations $\mean{\vec e\cdot\vec F}_{\x}$. The origin of the second contribution is the local detailed balance condition, which reduces $k^\pm$ if the system goes up in potential energy. The first contribution can then be interpreted as a constant ``housekeeping'' dissipation to keep the system away from equilibrium, which is modified by interactions. The strength of the second contribution, however, is not related to properties of the active particles but to the underlying propulsion mechanism and the solvent through $D^\text{c} = k_\text{eff}\lam^2$. Recall that $\keff$ depends on the total concentration of solute molecules [Eq.~\eqref{eq:keff}]. In equilibrium ($v_0 = 0$ and $\Fex=0$) the solute flux vanishes $\mathcal J=0$ since  $\mean{\vec e\cdot\vec F}_{\x}=0$.

In the linear response regime, i.e., if the system operates close to thermal equilibrium ($\beta\Delta\mu\ll 1$), the self-propulsion speed reduces to
\begin{equation}
  v_0 = k_\text{eff} \lambda \beta \Delta \mu.
  \label{eq:LR_speed_mob}
\end{equation}
The bulk dissipation rate on the mesoscale [Eq.~\eqref{eq:epr:ABPs}] to linear order in $\Delta\mu$ (and in the absence of external forces) reads
\begin{equation}
  \dot Q^\text{lin} = \sum_{i=1}^N \Int{\vec{r}^N \dd \vec{e}^N} \left[ \frac{v_0^2}{\mu^\text{c}}  - v_0 \vec{e}_i\cdot (\nabla_i U)  \right] \psi
  \label{eq:epr:lin}
\end{equation}
when expressed as a function of the propulsion speed $v_0$. This result has been obtained previously by Pietzonka and Seifert for the dissipation rate of catalytic particles through examining the continuum limit of a lattice model~\cite{pietzonka18}. The fact that it recovers the dissipation only to linear order in $\Delta\mu$ is caused by the necessity of taking the $\Delta \mu \to 0$ limit when coarsening the dynamics to a continuous state space. We emphasize that in deriving Eq.~\eqref{eq:epr:ABPs}, and therefore also the macroscopic dissipation rate [Eq.~\eqref{eq:epr:field}], no such assumption is made and it holds arbitrarily far from equilibrium.

\subsection{Linear irreversible thermodynamics}
\label{sec:LIT}

The linear response regime close to thermal equilibrium ($\Delta\mu=0$) is of particular interest since the dissipation rate should follow from more generic arguments that are agnostic towards microscopic details. The framework of linear irreversible thermodynamics requires the identification of driving affinities and their conjugate macroscopic currents which describe the exchange between system and reservoirs~\cite{onsager31, onsager31a, groot84}. This framework has been applied to active matter systems~\cite{markovich21,gaspard17,gaspard18, gaspard19, julicher18}. A major advantage of our bottom-up approach is that we can now demonstrate how such a framework unfolds.

In what follows, we consider the macroscopic thermodynamic affinities $\beta\Delta\mu$ and $\beta\Fex$ as small perturbations away from thermal equilibrium. As a first step, we seek closed expressions, to linear order in the affinities, for the force average $\mean{\vec F}_{\vec r}$ and the interaction-orientation correlation function $\mean{\vec e \cdot \vec F}_{\vec r}$ that appear in Eqs.~\eqref{eq:cur_p_macro} and \eqref{eq:cur_c_macro}, respectively. A linear response calculation yields (for details see Appendix~\ref{app:LR})
\begin{equation}
  \mean{\vec F}_{\vec r} = \mean{\vec F}^{\text{eq}}_{\vec r} \quad \text{and} \quad  \mean{\vec e \cdot \vec F}_{\vec r} = 0.
  \label{eq:mean_corr}
\end{equation}
Here we denote averages with respect to the equilibrium distribution $\psi^\text{eq} \sim e^{-\beta U}$ as $\mean{\cdot}^\text{eq}$ and use the fact that in equilibrium: (i) force averages vanish, and (ii) $\mean{\vec e \cdot \vec F}^{\text{eq}}_{\vec r} = 0$ since orientations and interactions decorrelate. Remarkably, pair interactions completely decouple from thermodynamic affinities at linear order. Moreover, in homogeneous equilibrium states the interaction contributions vanish all together.

By inserting Eq.~\eqref{eq:LR_speed_mob} and collecting terms, one can express the coupling between particle current [Eq.~\eqref{eq:cur_p_macro}] and chemical current [Eq.~\eqref{eq:cur_c_macro}] via a symmetric Onsager matrix
\begin{align}
  \begin{pmatrix} \vec{j}+\mu_0\nabla\cdot\msig \\ j^\text{c}+\keff\lam\nabla\cdot\vec{p} \end{pmatrix} 
  = \begin{pmatrix} D_0\rho \id & k_\text{eff} \lambda \vec{p} \\ k_\text{eff} \lambda \vec{p}^\intercal & k_\text{eff} \rho \end{pmatrix} \begin{pmatrix}  \beta\Fex \\ \beta \Delta \mu \end{pmatrix}.
  \label{eq:currents_LIT}
\end{align}
Note that both currents split into a ``thermodynamic'' contributions on the right and a kinetic contribution on the left hand side. For the particle current, the latter can be represented as the divergence of the symmetric stress tensor $\msig = -\kT \rho \id + \msig_\text{IK}$, where we identify the Irving-Kirkwood stress tensor $\msig_\text{IK}$ through $\nabla \cdot \msig_\text{IK} = \mean{\vec F}^{\text{eq}}_{\vec r}$~\cite{irving50}. We emphasize that no assumption has been made that restricts the applicability of the result to homogeneous states, since $\msig$ can be arbitrarily large. Furthermore, polarization fields appearing in Eq.~\eqref{eq:currents_LIT} should be evaluated in equilibrium and consequently vanish. Thus, particle and chemical currents completely decouple for linear perturbations around thermal equilibrium. This no longer applies for linear perturbations around \emph{non-equilibrium stationary states}. As we show in Appendix~\ref{app:LIT_NESS}, the coupling between affinities and currents is akin to Eq.~\eqref{eq:currents_LIT} but with now non-zero and no longer symmetric off-diagonal elements. Furthermore, both currents attain an additional ``housekeeping'' contribution necessary to maintain the non-equilibrium dynamics.

Within linear irreversible thermodynamics~\cite{groot84}, the local dissipation rate follows as the sum of products between linear Onsager currents [left hand side of Eq.~\eqref{eq:currents_LIT}] and the associated generalized forces $\{\beta \Fex, \beta \Delta\mu \}$,
\begin{align}
  \dot q^\text{lin} = \beta \left( \vec{j}+\mu_0\nabla\cdot\msig \right) \cdot \Fex + j^\text{c}\beta \Delta \mu. 
  \label{eq:lepr:lin}
\end{align}
This suggests a contribution of the particle current to the local dissipation rate in contrast to the identification in Eq.~\eqref{eq:lepr:fields}. To further investigate, we express the first term on the right hand side as the divergence of a gauge field by exploiting the stationary state continuity equation $\nabla \cdot \vec j = 0$ to write $j_k = \nabla \cdot (\vec j x_k)$ for the $k$th component of the particle current. Subsequent use of the divergence theorem leads to the 
integral over the subsystem boundary $\partial \Omega$
\begin{multline}
    \IInt{\vec r}{\Omega}{} \beta \left( \vec{j}+\mu_0\nabla\cdot\msig \right) \cdot \Fex \\ = \OInt{A}{\partial \Omega}{} \beta \left[(\vec n \cdot \vec j)\vec r + \mu_0 \vec n \cdot \msig \right] \cdot \Fex.
    \label{eq:lit_current}
\end{multline}
Since this expression is generally non-zero (for arbitrary choices of $\Omega$), Eq.~\eqref{eq:lepr:lin} contradicts the exact result obtained in Eq.~\eqref{eq:lepr:fields}. This implies that naively employing the framework of irreversible thermodynamics fails to correctly predict the spatially-resolved dissipation rate. Nevertheless, if $\Omega$ extends the full system the gauge field in Eq.~\eqref{eq:lit_current} vanishes for the considered boundary conditions (cf.~Appendix~\ref{app:BC}) and the global dissipation rate in linear response amounts to 
\begin{equation}
  \dot Q^\text{lin} = \Int{\vec r} j^\text{c} \Delta \mu 
  = \Int{\vec r} \frac{v_0^2}{\mu^\text{c}} \rho.
  \label{eq:dissipation_LIT}
\end{equation}
In agreement with Eq.~\eqref{eq:epr:field}, it is solely determined by chemical events and does not involve the diffusive particle current. Additionally, since interactions and orientations are uncorrelated at linear order, the only non-zero contribution stems from the continual dissipation that is necessary to maintain directed motion. Thus, we conclude that in our case the phenomenological theory of irreversible thermodynamics correctly captures the global dissipation, but fails to identify the appropriate local contributions. While it is tempting to conjecture this to be a consequence of the local non-equilibrium nature of active systems, a general statement requires further examination.

\subsection{Illustrations}
\label{sec:illustrations}

\subsubsection{Flat wall}
\label{sec:wall}

\begin{figure*}
  \includegraphics[scale=1]{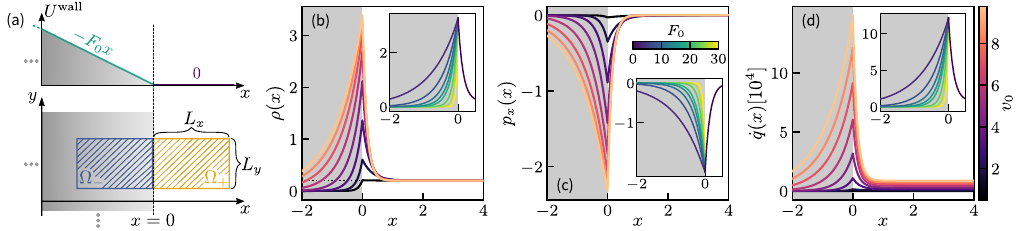}
  \caption{Catalytic particles in the presence of a straight potential wall. (a)~Sketch of the considered system, where for $x<0$ (shaded gray regions) particles experience a constant force due to the wall potential $U^\text{wall} = -F_0 x$ and are freely moving for $x>0$. The bottom axis shows subsystems $\Omega_\pm$ of same size $L_x \times L_y$, sharing a boundary at $x=0$. (b)~Density profiles $\rho(x)$, (c)~polarization profiles $p_x(x)$, and (d)~the local dissipation rate $\dot q(x)$ for $F_0=8$. Colors correspond to different self-propulsion speeds $v_0$ obtained by varying $\Delta \mu$ (and setting $\lam = 10^{-3}$ with $\keff \lam = 5$). Insets depict the respective profiles for a fixed value of $v_0 \approx 9.6$ with varying $F_0$, as indicated by the colors. The dotted black line in panel (b) shows the bulk density $\rho_\text{b} = 0.2$.}
  \label{fig:wall}
\end{figure*}

To get a better understanding of Eq.~\eqref{eq:epr:field} and its components [Eqs.~\eqref{eq:flux:field} and~\eqref{eq:boundary}], it is instructive to consider a simple, analytically solvable example and explicitly calculate its dissipation rate. To this end, we consider a two-dimensional system in the presence of a flat wall positioned at $x=0$ [see Fig.~\ref{fig:wall}(a)]. For simplicity, we assume non-interacting particles and close the hierarchy of hydrodynamic equations at the nematic order by setting $\vec Q \approx 0$. Throughout the discussion, we report values of quantities as dimensionless by measuring lengths in units of particle diameters $\sig$, energies in units of $\kT$ and times in units of $\sig^2/D_0$. Moreover, we set the rotational diffusion coefficient to $D_\text{r} = 3 D_0/\sig^2$, as is customary for hard disks due to the no-slip boundary of colloidal particles~\cite{buttinoni13,bialke15}. For equations we retain proper dimensions. 

In the $x>0$ region, particles are freely self-propelling and experience no forces. For $x<0$ particles experience a constant force of strength $F_0$ directed along $x$ due to the linear potential $U^\text{wall} = -F_0 x$. Hence, the limit $F_0 \to \infty$ results in a hard wall. Note that by construction the system is symmetric along $y$, allowing us to write density $\rho(x)$ and non-zero polarization component $p_x(x)$ as functions of $x$ only. The solutions are derived in Appendix~\ref{app:wall} and given by Eqs.~\eqref{app_eq:den_gen} and \eqref{app_eq:pol_gen}, with coefficients in Eq.~\eqref{app_eq:wall_coeff}. A visual representation is provided by Fig.~\ref{fig:wall}(b) and (c) for different self-propulsion speeds and force strengths (see insets). The results agree with the known trend that motile particles have the tendency to aggregate at walls with polar alignment perpendicular to the boundary~\cite{yan15, speck16a, duzgun18}. Upon increasing the force strength, exploration of the negative half-space is suppressed, leading to steeper exponential decays in both density and associated polarization. 

Using Eq.~\eqref{eq:lepr:fields}, we calculate the local dissipation rate [see Fig.~\ref{fig:wall}(d)], which upon comparison with Fig.~\ref{fig:wall}(b) clearly shares the characteristics of the density profile as expected. We emphasize that the exact values displayed in Fig.~\ref{fig:wall}(d) are highly sensitive to the choice of $\keff$ and $\lam$. Since the self-propulsion speed $v_0$ [Eq.~\eqref{eq:v0}] is restricted to values in the interval $v_0\in [-2\keff \lam, 2\keff \lam]$, due to the bounded hyperbolic tangent, the range of attainable speeds is determined by the product $\keff \lam$. For instance, by fixing $\keff \lam = 5$ as is done for Fig.~\ref{fig:wall}, speeds saturate at a value of $v_0 =10$. Moreover, by fixing $\keff \lam$ the magnitude of the density-dependent term scales inversely with $\lam$, i.e., if chemically induced displacements are short, maintaining a speed $v_0$ requires additional reactions which further promotes the dominance of the density-dependent source term in Eq.~\eqref{eq:cur_c_macro}. Nonetheless, the discussed trend is representative for physically sensible parameter choices.

Near the wall, locally induced polar order contributes to the dissipation rate in two ways. First, for $x>0$ ($x<0$) the positive (negative) divergence of polarization $\vec p$ [see last term in Eq.~\eqref{eq:cur_c_macro}] reduces (enhances) dissipation. Second, the strong anti-alignment between forces acting in the negative half-space and the polarization field lead to an additional negative contribution for $x<0$. This demonstrates that boundaries and obstacles play highly non-trivial roles in shaping the system's dissipation. Regardless, the dominant source term overshadows these intricacies due to the greatly increased local density by particles aggregating at the wall.

Note that the dissipation rate is well-defined anywhere but $x=0$ due to the cusp in the polarization field. Both limits $\dot q(x\to \pm 0)$, however, exist. For $x \to - \infty$, dissipation vanishes due to the absence of particles and for $x \to \infty$ it converges to the constant bulk value $\dot q(x\to \infty) = v_0\rho_\text{b} \Delta \mu/\lam$.

We now shift our attention to the finite subsystems $\Omega_\pm$ of area $L_x \times L_y$ that share a boundary at $x = 0$, as depicted in Fig.~\ref{fig:wall}(a), and calculate the respective average total dissipation rates $\dot Q_\pm = (\mathcal{J}_\pm - \dot \Gamma_\pm) \Delta \mu$. Assuming sufficiently large $L_x$, such that $\rho(L_x) \approx \rho_\text{b}$ and $\rho(-L_x) \approx 0$, the only contribution to the contour integral in Eq.~\eqref{eq:boundary} stems from the shared boundary at $x=0$ since polar order vanishes in the bulk and $p_y=0$. Consequently, for $x>0$ one obtains Eq.~\eqref{app_eq:flux_plus} for the average solute flux and
\begin{align}
  \dot \Gamma_+ = -\frac{D^\text{c}}{\lam} \IInt{y}{0}{L_y} p_x(0) = \frac{D^\text{c}}{\lam}\frac{D_0 C_+ }{v_0 \xi} L_y
\end{align}
for the boundary term with decay length $\xi \equiv \sqrt{D_0 / D_\text{r}} / \sqrt{1+ v_0^2/(2D_\text{r}D_0)}$ and coefficient $C_+ = \rho_\text{b} v_0^2/(2 D_\text{r}D_0)$. 
Similarly, for $x<0$ we find the solute flux [Eq.~\eqref{app_eq:flux_minus}]
and the boundary contribution
\begin{align}
  \dot \Gamma_- = -\frac{D^\text{c}}{\lam}\frac{D_0 C_+}{v_0 \xi} L_y = -\dot \Gamma_+.
\end{align}
Notably, all contributions individually vanish in equilibrium ($\Delta \mu=0$).
Moreover, $\dot \Gamma_-$ is given by its positive half-space counterpart $\dot \Gamma_+$ under change of sign, indicating the exchange of heat between the two subsystems. More precisely, in this example $\Omega_+$ transfers heat with rate $\dot Q^\text{tr}_+ = \dot \Gamma_+ \Delta \mu$---that was originally extracted in form of chemical work from the chemostat---to the neighboring subsystem $\Omega_-$, mediated by the non-zero polarization at the shared boundary along $x=0$. Consequently, obstacles or confinements that induce local polar order can, depending on the specific choice of subsystem, take the role of either a dissipation sink ($\Omega_+$) or source ($\Omega_-$). Crucially, the energy of the composite system is still conserved, as these effects vanish for bulk averages. This can be seen by, e.g., considering the combined system $\Omega = \Omega_+ + \Omega_-$. For the combined energy balance boundary contributions cancel each other, resulting in the first law $\dot Q_\Omega = (\mathcal{J}_+ + \mathcal{J}_-)\Delta\mu = \mathcal W^\text{in}_\Omega$, which assures the conservation of energy by equating the rates of heat dissipation and work injection.

\subsubsection{Motility-induced phase separation---MIPS}
\label{sec:MIPS}

\begin{figure*}
  \includegraphics[scale=1]{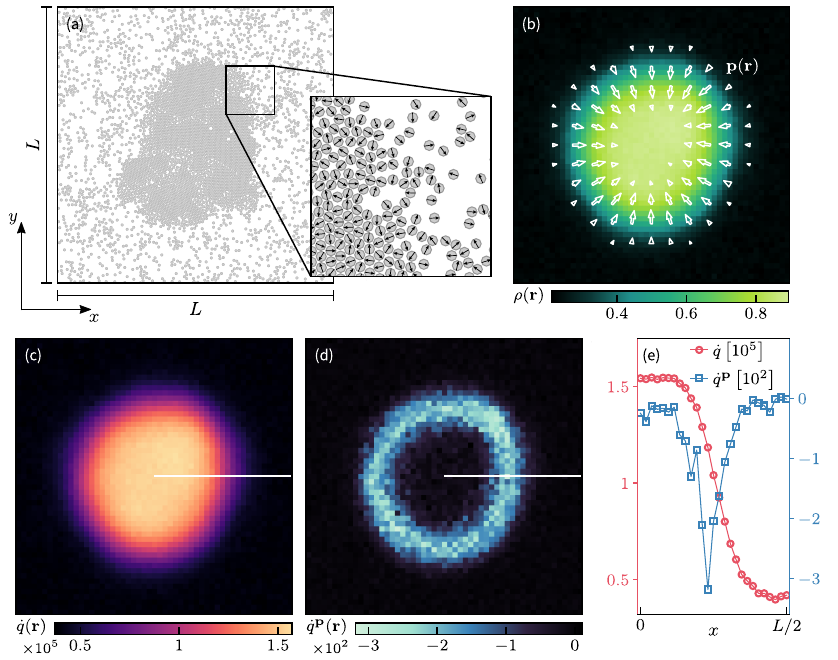}
  \caption{Suspension of active Brownian particles undergoing MIPS. (a)~Simulation snapshot showing the coexistence of a dense liquid cluster surrounded by a gaseous background in an $L \times L $ square [for details see Appendix~\ref{app:sim_MIPS}]. The inset shows a zoom-in on the marked region, where arrows indicate individual particle orientations. (b)~Density field $\rho(\vec r)$ (colors) and polarization field $\vec p(\vec r)$ (arrows), obtained by averaging over binned simulation data. Arrow sizes indicate the polarization magnitude. (c)~Local dissipation rate $\dot q(\vec r)$ calculated according to Eq.~\eqref{eq:lepr:fields} and (d)~the dissipation rate without density-dependent source terms, $\dot q^{\vec{p}} = \dot q- v_0\Delta \mu \rho/\lam$, using the fields depicted in panel (b). (e)~Profiles of local (circles) and residual (squares) dissipation rates, showing a cross section of the system as indicated by the white lines in panels (c,\,d). Lines connecting markers are a guide to the eye. Here we use parameter values: $v_0=80$, $\lam = 10^{-3}$ and fixed the product $\keff \lam = 50$.}
  \label{fig:MIPS}
\end{figure*}

As a second example, we consider a suspension of interacting motile particles that exhibits MIPS~\cite{buttinoni13, palacci13, cates15} and compute its spatially resolved dissipation rate. This allows us to uncover the mechanisms contributing to heat dissipation and characterize the role of local energy consumption to maintaining the non-equilibrium phase coexistence.

We analyze simulation data of $N=4000$ active colloids in a periodic $L\times L$ square box, with $L\approx102.3$ and a packing fraction of $\phi=0.3$. Particles interact via a repulsive Weeks-Chandler-Anderson potential and we apply no external drift, $\Fex=0$ (for details see Appendix~\ref{app:sim_MIPS}). Measuring quantities in the same units as in the proceeding section (Sec.~\ref{sec:wall}), particles are self-propelled with speed $v_0 = 80$ and consequently undergo MIPS as predicted from the well-explored phase diagram of ABPs~\cite{bialke15, siebert18, klamser18}. A simulation snapshot of the phase coexistence is shown in Fig.~\ref{fig:MIPS}(a). By averaging binned data over several time steps, we obtain the coarse grained hydrodynamic fields depicted in Fig.~\ref{fig:MIPS}(b), where colors indicate the density and arrows the polarization field. Arrows are scaled proportional to the polarization magnitude and indicate polar ordering at the liquid-gas interface.

Using Eq.~\eqref{eq:lepr:fields}, the local dissipation rate amounts to the values displayed in Fig.~\ref{fig:MIPS}(c) and (d). Here we used $\lam=10^{-3}$, set $\keff \lam=50$, and emphasize that the following discussion is representative for physically sensible parameter choices, namely $\lam \ll \sig$. In analogy with the results obtained in Sec.~\ref{sec:wall}, heat dissipation is largely determined by the density-dependent housekeeping contribution, i.e., the first term of Eq.~\eqref{eq:cur_c_macro} that enters Eq.~\eqref{eq:lepr:fields}. We emphasize that this term is not conditioned on particle displacements per se, which means that although active translations are inhibited, colloids arrested in the dense cluster maintain persistent fuel consumption. Consequently, the local dissipation is maximal within the liquid domain.

In close vicinity of the interface, the strong anti-correlation between particle orientations (pointing towards the cluster center) and repulsive interaction forces (pointing outwards), i.e., $\mean{\vec e \cdot \vec F}_{\vec r} < 0$, decreases the chemical current. If one corrects for the housekeeping term the residual heat rate $\dot q^{\vec{p}} \equiv \dot q- v_0\Delta \mu\rho/\lam$ forms a ring along the cluster boundary, taking negative values only [see Fig.~\ref{fig:MIPS}(d,\,e)]. Hence, the interplay between local polar order and interactions reduces the amount of work required to maintain the non-equilibrium phase separation. Nevertheless, this correction is orders of magnitude smaller than the dominant contribution $v_0\Delta \mu\rho/\lam$ [cf.~Fig~\ref{fig:MIPS}(c-e)]. 

In the gaseous phase, interaction forces and polarization density are negligibly small and dissipation is decided by the corresponding bulk density alone. As a result, the work extracted from the chemostat reduces drastically and converges to a non-zero bulk value, preserving particle motility and keeping the system out of (local) equilibrium. 

Complementary studies concerning the informatic entropy production of a scalar active field theory undergoing MIPS (active model B~\cite{wittkowski14}), in contrast report a strongly peaked irreversibility measure at the liquid-gas interface and small constant values within the bulk phases~\cite{nardini17}. This should not come as a surprise since path entropy can only account for the irreversibility encoded in the effective model dynamics, which manifests through density gradients at the interface and is agnostic to the energy dissipated by the underlying self-propulsion mechanism. Therefore, even in the dense region where heat dissipation and consequently the distance to a state of thermal equilibrium is greatest, it returns only minimal values. By subtracting the density-dependent source term entering Eq.~\eqref{eq:lepr:fields}, our local dissipation rate shows similar characteristics [see Fig.~\ref{fig:MIPS}(d,\,e)]. Crucially, however, although maintenance of the interface between coexisting phases is stimulated by the non-equilibrium nature of the dynamics, dissipation itself is reduced due to the polar order. Hence, knowledge of the heat dissipation in active systems (and by association the \emph{thermodynamic} entropy production) requires the rigorous bookkeeping of the partaking degrees of freedom and cannot be inferred solely from the resulting effective model dynamics and its time-reverse.

\subsection{Scalar field theory}
\label{sec:scalar}

Before concluding, we briefly sketch what dissipation can be inferred when going the final step to a scalar field theory through (adiabatically) eliminating the polarization field and all higher moments. To do so, we consider the limit of large length scales and time scales that are considerably longer than the characteristic decay time of orientational moments $D^{-1}_\text{r}$~\cite{cates13, bialke13}. Moreover, we set $\Fex = 0$ and employ a force closure to the conditional force to truncate the hierarchy~\cite{bialke13}. The latter allows us to rewrite Eq.~\eqref{eq:force_cond} as $\vec F = -\zeta \rho \vec e$, where the coefficient $\zeta$ depends on the pair potential $u(r)$ and the pair distribution $g(\vec r^\prime - \vec r | \vec e) = \psi_2 / \rho$. With these assumptions, we solve Eq.~\eqref{eq:eom_pol} for the polarization field
\begin{align}
    \vec p^\text{ad} \approx -\frac{v_0 - 2\mu_0 \zeta \rho}{2D_\text{r}} \nabla \rho,
    \label{eq:p_adia}
\end{align}
which thus adiabatically follows changes of the particle density. The local density is the sole scalar order parameter of the theory.

Plugging Eq.~\eqref{eq:p_adia} together with $\mean{\vec e\cdot\vec F}= -\zeta \rho$ for the correlations into the expression of the local dissipated heat rate [Eq.~\eqref{eq:lepr:fields}], we find
\begin{align}
    \dot q^\text{ad}(\vec r) =\Delta \mu \left[\frac{v_0}{\lam}\rho - \zeta\frac{\mu^\text{c}}{\lam}\left( \rho^2 + \frac{D_0}{2D_\text{r}} \nabla^2 \rho^2  \right)\right] 
\end{align}
to linear order in $\lam$. The first term again captures the incessant energy consumption required to drive the directed motion and to keep the system from reaching thermal equilibrium. In contrast to the general result in Eq.~\eqref{eq:lepr:fields}, the second term suggests a density (and spatial derivatives thereof) dependent reduction of dissipation in the adiabatic approximation. Hence, in this simplest of cases, divergence of the polarization is substituted by a term of order $\mathcal{O}(\rho^2, \nabla^2)$ and the orientation-interaction correlation function simply reduces to the squared local density. 

While higher-order expansion schemes have recently been shown to recover the popular extensions of classical model B~\cite{hohenberg77}, termed active model B(+)~\cite{wittkowski14, tjhung18}, an universally agreed upon recipe to derive microscopically-informed effective field parameters is still an open challenge~\cite{speck22, vrugt23, kalz23}. Nevertheless, as illustrated above, once connections between hydrodynamic fields and a single scalar order parameter are established, one can readily apply the derived expressions for the dissipation rate [Eqs.~\eqref{eq:epr:field} and \eqref{eq:lepr:fields}] to unravel thermodynamic details of the system at hand.

%% ---- conclusions ----

\section{Conclusions}
\label{sec:conclusion}

Over the last decade, active matter has emerged as a central paradigm for self-organization in soft and living matter, and has become a pillar of modern-day statistical physics. Concerned with the effects of local dissipation of residual or stored free energy, and its ramifications on the (dynamical) behavior of individual or large collectives of motile entities, active matter systems are by construction out of thermal equilibrium. This makes predicting physical properties a challenging endeavor, prompting immense efforts from experimentalists and theoreticians alike. Many models are inspired by biological systems, ranging from motile bacteria such as \emph{Escherichia coli}~\cite{berg72, berg04} to complex cellular reorganization during mitosis~\cite{needleman17, oriola18}. Any inquiry about the energetic cost to sustain cellular processes demands an accurate thermodynamic theory for active matter. One promising basis to build upon is the framework of stochastic thermodynamics, which extends notions of classical thermodynamics to microscopic systems governed by fluctuations. The use of stochastic thermodynamics requires the modeling of \emph{all} degrees of freedom contributing to the dissipation, and the few \emph{consistent} applications to active matter are usually restricted to single agents or the linear response regime~\cite{gaspard17, pietzonka18, speck18, gaspard18, gaspard19, fritz23, padmanabha23}. 

We have achieved the next step towards a thermodynamically consistent theory for active matter by deriving the exact heat dissipation rate of catalytically propelled colloidal particles, thereby bridging the gap between individual micron-sized particles and macroscopic collectives (Fig.~\ref{fig:scales}). By employing a systematic bottom-up approach, we have established connections between a microscopic model of a catalytic particle with explicit solute degrees of freedom, mesoscopic active Brownian particles, and macroscopic field theories. On the microscale, we have analytically shown how surface-mediated catalysis of molecular solutes leads to self-propulsion, without resorting to hydrodynamics. We derived the functional forms of the solute flux [Eq.~\eqref{eq:J-micro}] and the forces that fuel molecules exert on the colloidal particle [Eq.~\eqref{eq:force_res}], and showed that the former defines the particle self-propulsion speed together with an effective jump length $\lam$ [Eq.~\eqref{eq:lam}]. Microscopic details are condensed into two parameters: the effective attempt rate $\keff$ and the jump length $\lam$. To validate the theoretical prediction, we further performed Brownian dynamics simulations of a three-dimensional catalyst surrounded by explicit solute molecules. All results for the explicit model are faithfully reproduced by an effective model that tightly couples chemical events and particle translation. Introducing interactions between active particles, we arrived at a thermodynamically consistent many-body model that reduces to the well-known model of active Brownian particles [Eq.~\eqref{eq:abp}] in the (physically motivated) limit of small jump length $\lam$. We complement this model with the \emph{exact} expression for the dissipation [Eq.~\eqref{eq:epr:ABPs}]. From there we took the final step to coupled hydrodynamic equations for the evolution of (smooth) density and polarization fields. We note that our results extend to diffusiophoretic colloidal particles that are driven through the demixing of a binary solute~\cite{speck22a}.

The main benefit of our systematic bottom-up construction is that local thermodynamic information is preserved through every coarse-graining step, ultimately yielding exact expressions at the macroscale. Here we have focused on the dissipation and reported how tracking of the stochastic energetics allows the systematic identification of the solute flux $\mathcal{J}$ between chemical reservoirs at each level. This constitutes the injected work and determines both the dissipation rate $\Delta\mu\mathcal J_i$ and the speed $\lam\mathcal J_i$ of individual entities and macroscopic collectives. Such an exact expression for the dissipation also gives access to the linear-response regime and we demonstrated how the phenomenological framework of linear irreversible thermodynamics correctly predicts the global dissipation rate, but fails to capture local details that can be preserved through the bottom-up construction.

Beyond linear response, our results open new avenues to the design of active engines that exploit non-equilibrium fluctuations to extract work, by unveiling all mechanisms that constitute dissipation and the influence of external confinement.
Moreover, our general insights enable strategies to infer dissipation from observable large-scale data in a wide range of systems. For example, lower bounds to the actual dissipation provide valuable insights into living systems~\cite{seifert19, skinner21a, skinner21, daddi-moussa-ider23a, dechant23} but typically only capture a fraction of the actual dissipation. The identification of tightly-coupled degrees of freedom offers a new perspective on these systems.

%% ---- acknowledgments ----

\begin{acknowledgments}
  We acknowledge financial support by the Deutsche Forschungsgemeinschaft (DFG) through the collaborative research centers TRR 146 (grant no. 233630050) and SFB 1551 (grant no. 464588647). JFR gratefully acknowledges financial support from the Alexander von Humboldt foundation. We thank Sebastian Bauer for preliminary results on the explicit model (Sec.~\ref{sec:explicit}) and Ashreya Jayaram for code and data (Sec.~\ref{sec:MIPS}).
\end{acknowledgments}

%% ---- appendix ----

\appendix

\section{Explicit model}
\label{sec:thin}

\subsection{Thin interaction layer approximation}

Here we provide details for the calculation of the distribution of solute molecules around a single catalytic colloidal particle [cf.~Fig.~\ref{fig:sketch}(a)], and how this distribution determines fluxes and the force. To this end, we assume that there is a thin layer of width $\ell$ on the surface of the colloidal particle within which the densities of solutes quickly drop to zero as $r$ is decreased. In this appendix, we implicitly work in dimensionless quantities: length $\x\to\ell\x$, time $t\to k_0^{-1}t$ with rates $k^\pm(r,\theta)\to k_0h(\theta)k^\pm(r)$, and energies $u_\al\to\kT u_\al$. In steady state, the scaled current densities then obey
\begin{equation}
	\nabla\cdot\js_\al = \begin{cases}
		0 & (r>\rc) \\ \eps h(\theta)K_{\al \al^\prime} c_{\al^\prime} & (r<\rc)
	\end{cases}
	\label{eq:thin:divj}
\end{equation}
with $\js_\al(r,\theta)\equiv-u_\al'(r)c_\al\hat{\vec e}_r-\nabla c_\al$ and we have defined the small parameter $\eps\equiv \ell^2k_0/D\ll 1$. In this section we sum over repeated indices (summation convention). Continuity at $r = r_\text{C}$ is enforced by appropriate selection of coefficients. Note that here we neglect the diffusion of the catalytic particle (since $D_0\ll D$). The rate matrix reads
\begin{equation}
	\vec K(r) = \left(\begin{array}{rr}
		K_\text{SS} & K_\text{SP} \\ K_\text{PS} & K_\text{PP}
	\end{array}\right) = \left(\begin{array}{rr}
		-k^+ & k^- \\ k^+ & -k^-
	\end{array}\right)
	\label{eq:ratematrix}
\end{equation}
and $h(\theta)$ is an indicator function that is unity on active and zero on inert regions.

We expand the solute densities
\begin{equation}
	c_\al(r,\theta) = \sum_{l=0}^\infty a_{\al,l}(r)P_l(\cos\theta)
	\label{eq:legendre}
\end{equation}
in terms of Legendre polynomials $P_l(x)$ with coefficients $a_{\al,l}(r)$. For $\eps=0$, the full solution is the isotropic profile
\begin{equation}
	c_\al^{(0)}(r) = a^{(0)}_{\al,0}(r) = \bar c_\al e^{-u_\al(r)}, \quad
  a^{(0)}_{\al,l>0} = 0
  \label{eq:c0}
\end{equation}
with constant densities $\bar c_\al$ outside the interaction range, which are determined by the chemostat.
Note that Eq.~\eqref{eq:c0} is simply the equilibrium profile of an ideal gas in an external field (i.e., the barometric law).  
In the free region $r>\rc$ (superscript $>$) with boundary conditions $c_\al(r\to\infty)=\bar c_\al$, we have $a^>_{\al,l}(r)=\bar c_\al \delta_{l0}+b_{\al,l}r^{-(l+1)}$ with constant coefficients $b_{\al,l}$.

Within the interaction region ($r<\rc$), we expand $c_\al=c_\al^{(0)}+\eps c_\al^{(1)}+\mathcal{O}(\eps^2)$. Projecting Eq.~\eqref{eq:thin:divj} on the $l$th Legendre polynomial within a sphere of radius $r$ leads to
\begin{equation}
	\frac{2l+1}{2}\IInt{^3\x}{|\x|<r}{}P_l(\cos\theta)\nabla\cdot\js_\al^{(1)} = h_lI_\al(r)
	\label{eq:proj}
\end{equation}
with coefficients
\begin{equation}
  h_l \equiv \frac{2l+1}{2}\IInt{\theta}{0}{\pi}\sin\theta h(\theta)P_l(\cos\theta)
  \label{eq:h_l}
\end{equation}
giving $h_0= 1/2$ and $h_1=3/4$, and integral
\begin{align}
	I_\al(r) &\equiv 2\pi\IInt{\xi}{0}{r} \xi^2 \sum_{\al^\prime} K_{\al \al^\prime}(\xi)\bar c_{\al^\prime} e^{-u_{\al^\prime}(\xi)} \nonumber
  \\ 
  &= -2\pi(\bar c_\text{S}-\bar c_\text{P})\IInt{\xi}{0}{r} \xi^2 k_\al(\xi)e^{-u_\al(\xi)}
  \label{eq:I-integral}
\end{align}
involving the zero-order solution $c_\al^{(0)}$. The second line is obtained through utilizing the detailed balance condition [Eq.~\eqref{eq:db}], defining $k_\text{S}\equiv k^+$ and $k_\text{P}\equiv-k^-$ along the way. Due to the shape of the rate matrix [Eq.~\eqref{eq:ratematrix}], we immediately see that $I_\text{S}=-I_\text{P}$, which is a consequence of mass conservation.

To rewrite Eq.~\eqref{eq:proj}, we exploit $P_l\nabla\cdot\js_\al=\nabla\cdot(P_l\js_\al)-\js_\al\cdot\nabla P_l$ together with the divergence theorem for the first term to obtain the integro-differential equation
\begin{multline}
	-2\pi r^2\left\{u'_\al(r)a^{(1)}_{\al,l}(r)+[a^{(1)}_{\al,l}(r)]'\right\} \\ + 2\pi \sum_{k=0}^\infty \IInt{\xi}{0}{r} M_{lk}a^{(1)}_{\al,k}(\xi) = h_lI_\al(r)
	\label{eq:proj2}
\end{multline}
for the coefficients $a^{(1)}_{\al,l}$. We have defined
\begin{equation}
	M_{lk} \equiv \frac{2l+1}{2}\IInt{x}{-1}{1}(1-x^2)P'_k(x)P_l'(x) = l(l+1)\delta_{lk},
\end{equation}
with the final expression following after integrating by parts and using Legendre's differential equation.

\subsection{Solute flux}

The first quantity we are interested in is the flux between the reservoirs. Plugging the series Eq.~\eqref{eq:legendre} into the expression
\begin{align}
	J_\al(r) &= 2\pi r^2\eps^{-1}\IInt{\theta}{0}{\pi}\sin\theta \hat{\vec e}_r\cdot\js_\al 
  \nonumber \\ 
  &= -4\pi r^2\eps^{-1}[u_\al'(r)a_{\al,0}(r)+a'_{\al,0}(r)]
	\label{eq:flux}
\end{align}
for the flux of molecular solutes through a spherical surface with radius $r$ leads to an expression that depends only on the $l=0$ coefficient $a_{\al,0}(r)$. In the free region, $J^>_\al=4\pi\eps^{-1}b_{\al,0}$ is manifestly independent of $r$. Inside the interaction range, we see that the square brackets in Eq.~\eqref{eq:flux} are zero for $a_{\al,0}^{(0)}=c_\al^{(0)}$ and we use Eq.~\eqref{eq:proj2} to obtain $J^{(1)}_\al(r)=2h_0I_\al(r)$ at the next order. Since $I_\text{S}=-I_\text{P}$ we find $J_\text{S}(r)+J_\text{P}(r)=0$ everywhere, which guarantees the conservation of the total number of molecular solutes. The difference
\begin{equation}
  \mathcal J\equiv J^{(1)}_\text{P}(\rc)-J^{(1)}_\text{S}(\rc)=4h_0I_\text{P}(\rc)
  \label{eq:flux:net}
\end{equation}
is the net flux from the substrate reservoir to the product reservoir. Inserting $h_0 =1/2$ and Eq.~\eqref{eq:I-integral} into this expression gives Eqs.~\eqref{eq:J-micro} and \eqref{eq:keff} in the main text.

\subsection{Force}

For the propulsion speed we need the forces exerted by the solute molecules on the colloidal particle, which read 
\begin{align}
	\vec g_\al &= \IInt{r}{0}{\rc}r^2\IInt{\theta}{0}{\pi}\sin\theta\IInt{\vhi}{0}{2\pi} [u_\al'(r)\hat{\vec e}_r]c_\al(r,\theta) 
  \nonumber \\ 
	&= \frac{4\pi}{3}\IInt{r}{0}{\rc} r^2u_\al'(r)a_{\al,1}(r) \hat{\vec e}_z
  \label{eq:force}
\end{align}
with total force $\vec g=\vec g_\text{S}+\vec g_\text{P}$.
The force thus requires knowledge of the function $a_{\al,1}^{(1)}(r)$. Taking the derivative of Eq.~\eqref{eq:proj2} with respect to $r$, dividing by $2\pi r^2$, and writing $a_{\al,1}^{(1)}=\chi_\al e^{-u_\al}$ leads to the linear differential equation
\begin{equation}
	\nabla^2\chi_\al - \frac{2}{r^2}\chi_\al - u_\al'\chi_\al' = h_1(\bar c_\text{S}-\bar c_\text{P})k_\al
\end{equation}
for $\chi_\al(r)$. Since the homogeneous solution has to be zero [cf.~Eq.~\eqref{eq:c0}], we are only interested in the particular solution $\chi_\al(r)=h_1(\bar c_\text{S}-\bar c_\text{P})\Int{\bm\xi}G_\al(|\x-\bm\xi|)k_\al(\bm\xi)$. The (unknown) Green's function $G_\al(r)$ depends on the pair potential $u_\al(r)$. Inserting this solution into Eq.~\eqref{eq:force} and restoring dimensionful quantities leads to the force given in Eq.~\eqref{eq:force_res} of the main text, with a functional
\begin{equation}
  \mathcal V \equiv \frac{4\pi}{3} \sum_{\al} \IInt{r}{0}{r_\text{C}}r^2u_\al^\prime e^{-u_\al^\prime} \Int{\bm\xi} G_\al(|\vec r - \bm\xi|) k_\al(\bm\xi)
  \label{eq:V}
\end{equation}
that encodes the microscopic solute-colloid interactions but is independent of the chemostat.

\section{Choice of transition rates}
\label{app:rates}

In contrast to the exponential rates in Eq.~\eqref{eq:rates}, prevalently used in the field of stochastic thermodynamics, we briefly show that one could have similarly employed an ansatz based on the Glauber-rates in Eq.~\eqref{eq:glauber-rates}. Following a straight-forward generalization suggests rates of the form
\begin{align}
  k_i^\pm =k_\text{eff} \left[ 1+e^{\mp \beta \Delta\mu} e^{-\beta \left[- U(\vec{r}_i\pm\lam\vec{e}_i) + U(\vec{r}_i) + \lam \vec e_i \cdot \vec F^\text{ex} \right]} \right]^{-1}.
\end{align}
Clearly, local detailed balance is satisfied [cf.~Eq.~\eqref{eq:ldb}] and consequently thermodynamic consistency guaranteed.
After an expansion to linear order of small displacement length $\lam$ we obtain
\begin{align}
  k_i^\pm = \tilde{\kap}^\pm \left[1 \pm  \frac{\lam \beta }{e^{\pm\beta \Delta \mu}+1} \vec{e}_i \cdot(\vec F^\text{ex} -  \nabla_i U) \right] + \mathcal{O}(\lam^2),
\end{align}
with $\tilde{\kap}^\pm \equiv k_\text{eff} \left(1 + e^{\mp \beta \Delta \mu} \right)^{-1}$.
Plugging this expansion, alongside Eqs.~\eqref{eq:expans_psi} and \eqref{eq:expans_k}, into Eq.~\eqref{eq:diff_op} we recover an equation of motion identical to the one written in Eq.~\eqref{eq:abp} by defining 
\begin{equation}
  v_0 \equiv \lam (\tilde{\kap}^+ - \tilde{\kap}^-) \\
\end{equation}
and
\begin{equation}
  \mu^\text{c} \equiv \lam^2 \beta  \frac{2e^{\beta \Delta \mu}}{(e^{\beta \Delta \mu}+1)^2},
\end{equation}
as the corresponding speed and chemical mobility.

Lastly, in the linear response regime, assuming $\Delta \mu \beta \ll 1$, expansion to leading order results in $v_0=k_\text{eff}\lam\beta\Delta\mu/2$ and $\mu^\text{c}=k_\text{eff}\lam^2 \beta/2$, which after re-normalizing the attempt rate $k_\text{eff} \to 2k_\text{eff}$ recovers Eq.~\eqref{eq:LR_speed_mob}.

\section{Boundary terms}
\label{app:BC}

The multiplication of Eq.~\eqref{eq:smol_stat} by the total ``potential'' $U^\text{tot} = U - \sum_{i=1}^N \vec r_i \cdot \Fex $, including interaction and external forces, followed by repeated integration by parts leads to multiple surface terms. Retaining these surface terms, Eq.~\eqref{eq:condition} reads
\begin{widetext}
\begin{multline}
  \sum_{i=1}^N \Int{\vec{r}^N \dd \vec{e}^N} \left[ -v_0 \vec e_i\cdot (\nabla_i U^\text{tot})\psi + \mu_0 (\nabla_i U^\text{tot})^2 \psi +  D_0 (\nabla_i^2 U^\text{tot})\psi   \right] = \\ \sum_{i=1}^N \Int{\vec r^{N-1}\dd \vec e^N}\OInt{\vec r_i}{\partial V}{} \cdot \left[ v_0 \vec e_i U^\text{tot} \psi - \mu_0\psi U^\text{tot}\nabla_i U^\text{tot} - D_0 (U^\text{tot} \nabla_i \psi - \psi \nabla_i U^\text{tot})\right],
\end{multline}
\end{widetext}
where we denote the system boundary as $\partial V$. All surface terms on the right-hand side vanish provided $\psi\to0$ (holds for infinite systems and confining wall potentials) and for potentials $U^\text{tot}$ that are periodic and continuous across the boundaries.

We emphasize that any of the aforementioned assumptions is consistent with dropping the boundary terms in the total dissipation rate [Eq.~\eqref{eq:epr:ABPs}] and during the coarse-graining procedure presented in Sec.~\ref{sec:cg}.

\section{Linear response theory}
\subsection{Perturbations around equilibrium}
\label{app:LR}

We provide additional details on the linear response calculation preformed in Sec.~\ref{sec:LIT}. We start by considering the evolution equation of the joint probability distribution given in Eq.~\eqref{eq:abp} and write
\begin{align}
  \pd{\psi}{t} = (\mathcal{L}^{\text{eq}} + \mathcal{L}^{(1)} + \mathcal{L}^{(2)}) \psi,
  \label{app_eq:smol_split}
\end{align}
with the generator of the equilibrium dynamics 
\begin{align}
  \mathcal{L}^{\text{eq}} &\equiv \sum_{i=1}^N  \{ \nabla_i \cdot \left[\mu_0\nabla_i U+D_0\nabla_i  \right] - D_\text{r} \Delta_{\vec{e}_i} \}
\end{align}
and the associated generators of small perturbations 
\begin{align}
  \mathcal{L}^{(1)} &\equiv -\sum_{i=1}^N D_0 \beta \Fex \cdot \nabla_i \label{app_eq:gen_f} \\
  \mathcal{L}^{(2)} &\equiv -\sum_{i=1}^N \keff \lam \beta \Delta \mu \vec e_i \cdot \nabla_i.
  \label{app_eq:gen_mu}
\end{align}
In the linear regime, the general solution is formally given by
\begin{align}
  \psi = \psi^\text{eq} +\sum_{m \in\{1,2\}} \IInt{t^\prime}{-\infty}{t} e^{\mathcal{L}^\text{eq}(t-t^\prime)} \mathcal{L}^{(m)}\psi^\text{eq},
  \label{eq:LR_general}
\end{align}
with equilibrium distribution $\psi^\text{eq} \sim e^{-\beta U}$ and we denote equilibrium averages as $\mean{\cdot}^\text{eq}$. With the general solution [Eq.~\eqref{eq:LR_general}] the average of a coarse-grained observable $O$ (introduced as a proxy for $\vec F$ and $\vec e \cdot \vec F$) takes the form 
\begin{multline}
  \mean{O}_{\vec r} \approx \mean{O}_{\vec r}^\text{eq} + \sum_{m\in \{ 1,2 \}}\Int{\vec r^N \dd \vec e^N}\IInt{t^\prime}{-\infty}{t} \sum_{i=1}^N O_i \\ \times \delta(\vec r - \vec r_i) e^{\mathcal{L}^\text{eq}(t-t^\prime)} \mathcal{L}^{(m)}\psi^\text{eq}.
  \label{app_eq:observable_LR}
\end{multline}
Here we use the identity $\mean{O}_{\vec r} = \mean{\sum_{i=1}^N O_i \delta(\vec r - \vec r_i)}$, where $\mean{\cdot}$ labels the system average with respect to the perturbed distribution $\psi$, and $O_i$ the observable evaluated for individual particles $i$.
By letting the generators of the small perturbations $\mathcal L^{(1)}$ and $\mathcal L^{(2)}$ act on the equilibrium distribution, we find
\begin{align}
  \mathcal L^{(1)} \psi^\text{eq} &= \sum_{i=1}^N \mu_0 \beta (\Fex \cdot \nabla_i U) \psi^\text{eq} \\
  \mathcal L^{(2)} \psi^\text{eq} &= \sum_{i=1}^N \keff \lam \beta^2 \Delta \mu ( \vec e_i \cdot \nabla_i U) \psi^\text{eq}.
\end{align}
Plugging these into Eq.~\eqref{app_eq:observable_LR}, choosing $O \in \{ \vec F, \vec e \cdot \vec F\}$ and formally taking the $t \to \infty$ limit to obtain time-independent stationary expressions, the average 
\begin{multline}
  \mean{\vec F}_{\vec r} = \mean{\vec F}^\text{eq}_{\vec r} - \mu_0\beta \mean{\vec F}^\text{eq}_{\vec r} (\mean{\vec F}^\text{eq} \cdot \Fex)
  \\ 
  - \keff\lam\beta^2\Delta\mu \mean{\vec F}^\text{eq}_{\vec r} \mean{\vec e \cdot \vec F}^\text{eq} 
\end{multline}
and orientation-interaction correlation function 
\begin{multline}
  \mean{\vec e \cdot \vec F}_{\vec r} = \mean{\vec e \cdot \vec F}^\text{eq}_{\vec r} - \mu_0 \beta \mean{\vec e \cdot \vec F}^\text{eq}_{\vec r} (\mean{\vec F}^\text{eq} \cdot \Fex) 
  \\
  -\keff\lam\beta^2\Delta\mu \mean{\vec e \cdot \vec F}^\text{eq} \mean{\vec e \cdot \vec F}^\text{eq}_{\vec r}
\end{multline}
are expressed through equilibrium averages and correlation functions only. Lastly, realizing that orientations and interactions are independent in equilibrium and that the equilibrated state has to (on average) be force-free, we recover the results reported in Eq.~\eqref{eq:mean_corr}.

\subsection{Perturbations around a non-equilibrium stationary state}
\label{app:LIT_NESS}

To corroborate the discussion of equilibrium perturbations, we present here how non-equilibrium steady states respond to small changes $ \beta \delta\Delta \mu $ in the driving affinity $\beta \Delta \mu$ that maintains the distance to thermal equilibrium and weak external perturbations through applying force $\beta \Fex$. After perturbing $\Delta \mu \to \Delta \mu + \delta \Delta \mu$, we expand the self-propulsion speed [Eq.~\eqref{eq:v0}] in small $\delta \Delta \mu$ and define 
\begin{align}
    \bar v_0 &\equiv 2 \keff \lam \tanh\left(\frac{\beta \Delta \mu}{2}\right) \\
    \delta v_0 &\equiv \keff \lam \cosh^{-2}\left(\frac{\beta \Delta \mu}{2}\right), 
\end{align}
such that $v_0 \approx \bar v_0 + \delta v_0 \,\beta\delta\Delta \mu$.
Analog to Eq.~\eqref{app_eq:smol_split}, the Smoluchowski equation of joint distribution $\psi$, splits into three parts where the stationary state generator reads 
\begin{align}
  \mathcal{L}^{\text{s}} \equiv \sum_{i=1}^N  \left\{\nabla_i \cdot \left[\mu_0\nabla_i U + D_0\nabla_i - \bar v_0 \vec e_i \right]
  - D_\text{r} \Delta_{\vec e_i} \right\},
\end{align}
and perturbations are generated by operators 
\begin{align}
  \mathcal{L}^{(1)} &\equiv -\sum_{i=1}^N D_0 \beta \Fex \cdot \nabla_i \\
  \mathcal{L}^{(2)} &\equiv -\sum_{i=1}^N \delta v_0 \, \beta\delta \Delta \mu \,\vec e_i \cdot \nabla_i.
\end{align} 
Following the calculation presented in Appendix~\ref{app:LR}, the average interaction force and the orientation-interaction correlation function amount to 
\begin{align}
  \mean{\vec F}_{\vec r} = \mean{\vec F}^\text{s}_{\vec r} \quad \text{and} \quad  \mean{\vec e \cdot \vec F}_{\vec r} =  \mean{\vec e \cdot \vec F}^\text{s}_{\vec r},
\end{align}
since terms linear in the affinities cancel. Here we denote averages with respect to the (generally unknown) stationary measure $\psi^\text{s}$ as $\mean{\cdot}^\text{s}$. Plugging these results in the expressions for the particle and chemical current, given by Eqs.~\eqref{eq:cur_p_macro} and \eqref{eq:cur_c_macro} respectively, yields 
\begin{multline}
  \begin{pmatrix} \vec{j}+\mu_0 \nabla \cdot \msig^\text{s}  \\ j^\text{c} + \keff\lam \left(\nabla\cdot\vec{p} -\beta \mean{\vec e \cdot \vec F}^\text{s}_{\vec r} \right)\end{pmatrix} 
  =  \begin{pmatrix} \bar v_0\vec p  \\ \bar v_0 \rho/\lam\end{pmatrix} 
  \\ 
  + \begin{pmatrix} D_0\rho \id & \delta v_0 \vec{p} \\ k_\text{eff} \lambda \vec{p}^\intercal & \delta v_0 \rho/\lam \end{pmatrix} \begin{pmatrix}  \beta \Fex \\ \beta \delta \Delta \mu \end{pmatrix}.
\end{multline}
with stress tensor $\msig^\text{s} = -\kT \rho \id + \msig^\text{s}_\text{IK}$ and $\nabla \cdot \msig^\text{s}_\text{IK} = \mean{\vec F}^\text{s}_{\vec r}$. Notably, both the particle and chemical current attain an additional term that captures the incessant conversion of substrate molecules, required for housekeeping of the non-equilibrium steady state (first term on the right hand side). Moreover, the chemical current acquires an additional kinetic contribution due to the non-vanishing correlation function between orientations and interactions. 

\section{Active Brownian particles in presence of a flat wall}
\label{app:wall}

\subsection{Density and polarization profiles}

We provide details for the calculation of the density and polarization profiles displayed in Fig.~\ref{fig:wall}(b) and (c). A very similar problem was previously investigated in Refs.~\cite{hermann18, duzgun18}. We assume a two-dimensional system with an infinitely long straight wall, given by the linear potential $U^\text{wall} = -F_0 x$ for $x < 0$. The system is translationally invariant along $y$ allowing us to write the steady state evolution equations for density [Eq.~\eqref{eq:eom_density}] and polarization fields [Eq.~\eqref{eq:eom_pol}] as 
\begin{align}
  0 =& -v_0 \pd{p_x}{x} + D_0 \pd{^2\rho}{x^2} 
  \label{app_eq:den_free}\\
  0 =& -\frac{v_0}{2} \pd{\rho}{x} + D_0 \pd{^2 p_x}{x^2} - D_\text{r} p_x
  \label{app_eq:pol_free}
\end{align}
in the force-free $x>0$ region and 
\begin{align}
  0 =& -v_0 \pd{p_x}{x} - \mu_0 F_0 \pd{\rho}{x} + D_0 \pd{^2\rho}{x^2} 
  \label{app_eq:den_force}\\
  0 =& -\frac{v_0}{2} \pd{\rho}{x} -\mu_0 F_0 \pd{p_x}{x} + D_0 \pd{^2 p_x}{x^2} - D_\text{r} p_x
  \label{app_eq:pol_force}
\end{align}
under the influence of a constant force $F_0$ if $x<0$. Note that both $\rho$ and $p_x$ depend on $x$ only. To proceed, we assume that both fields take exponential forms ($\sim e^{\al_\pm x}$), with inverse decay lengths $\al_\pm$ for the positive and negative $x$-regions.

\paragraph*{$x>0$:} By rearranging Eqs.~\eqref{app_eq:den_free} and \eqref{app_eq:pol_free}, followed by insertion of the ansatz, one obtains the simple equation 
\begin{align}
  \al_+^3 = \frac{1}{\xi^{2}} \al_+,
\end{align}
with solutions $\al_+ = 0, \pm \xi^{-1}$, where we defined the decay length $\xi \equiv \sqrt{D_0 / D_\text{r}} / \sqrt{1+ v_0^2/(2D_\text{r}D_0)}$. Clearly, one can discard the positive solutions, since density and polarization must not diverge in the limit of large $x\to \infty$.

\paragraph*{$x<0$:} Along similar lines, rewriting Eqs.~\eqref{app_eq:den_force} and \eqref{app_eq:pol_force} results in 
\begin{align}
  \al_-^4 = 2\beta F_0 \al_-^3 + \left( \frac{1}{\xi^2} - \beta^2 F_0^2 \right) \al_-^2 - \frac{D_\text{r}}{D_0} \beta F_0 \al_-.
\end{align}
Assuming large forces $F_0 \gg 1$, the solutions to leading order in $F_0^{-1}$ read 
\begin{align}
  \al_- = 0,\quad -\frac{D_\text{r}}{\beta F_0 D_0},\quad \beta F_0 \pm \frac{v_0}{\sqrt{2}D_0}
\end{align}
and in analogy to aboves reasoning, we can immediately discard negative solutions as well as $\al_-=0$, due to the necessarily vanishing density and polarization as $x \to -\infty$.

Collecting terms, densities take the general form 
\begin{widetext}
\begin{align}
  \rho(x) = \begin{cases}
    C_+ e^{-x/ \xi} + B  &x>0 \\[1ex]
    C^{(1)}_{-} e^{\left[\beta F_0 + v_0/(\sqrt{2} D_0)\right]x} + C^{(2)}_{-} e^{\left[\beta F_0 - v_0/(\sqrt{2} D_0)\right]x} &x<0
  \end{cases}
  \label{app_eq:den_gen}
\end{align}
and polarization profiles are given by 
\begin{align}
  p_x(x) = \begin{cases}
    -\frac{D_0}{v_0 \xi} C_+ e^{-x/ \xi} &x>0 \\[1ex]
    \frac{1}{\sqrt{2}}\left(C_-^{(1)} e^{\left[\beta F_0 + v_0/(\sqrt{2} D_0)\right]x} - C_-^{(2)} e^{\left[\beta F_0 - v_0/(\sqrt{2} D_0)\right]x}\right) &x<0.
  \end{cases}
  \label{app_eq:pol_gen}
\end{align}
\end{widetext}
Note the necessity of sufficiently strong forces, such that $\beta F_0 > v_0/(\sqrt{2} D_0)$.
The remaining four coefficients are determined by enforcing that the density converges to the bulk value $\rho_\text{b}$ for $x\to \infty$, and that both density and polarization fields, as well as their respective currents, are continuous at $x=0$. Under these conditions the coefficients are given by
\begin{align}
  C_+ &= \frac{\rho_\text{b} v_0^2}{2 D_\text{r} D_0}, \qquad  B = \rho_\text{b}, \nonumber\\
  C_-^{(1)} &= \frac{1}{2} \left[ C_+ \left(1+ \frac{\sqrt{2} D_0}{v_0 \xi} \right) + \rho_\text{b}\right],
  \label{app_eq:wall_coeff}  \\
  C_-^{(2)} &= \frac{1}{2} \left[ C_+ \left(1 - \frac{\sqrt{2} D_0}{v_0 \xi} \right) + \rho_\text{b}\right], \nonumber
\end{align}  
which after insertion in Eqs.~\eqref{app_eq:den_gen} and \eqref{app_eq:pol_gen} give the solutions plotted in Fig.~\ref{fig:wall}(b) and (c) of the main text.

\subsection{Solute flux in subdomains}

With the analytical expressions for the density and polarization fields, given by Eqs.~\eqref{app_eq:den_gen} and~\eqref{app_eq:pol_gen} and the associated coefficients [Eq.~\eqref{app_eq:wall_coeff}], we calculate the average solute fluxes $\mathcal{J}_\pm$ for the two subdomains $\Omega_\pm$, depicted in Fig.~\ref{fig:wall}(a). 
Plugging the results into Eq.~\eqref{eq:flux:field} we immediately find 
\begin{align}
  \mathcal{J}_+ = L_y \IInt{x}{0}{L_x} \frac{v_0}{\lam} \rho = \frac{v_0}{\lam} L_y \left(L_x \rho_\text{b} + \xi C_+ \right)
  \label{app_eq:flux_plus}
\end{align}
for the solute flux in subregion $\Omega_+$ and similarly
\begin{align}
  &\mathcal{J}_- = L_y \IInt{x}{-L_x}{0} \left( \frac{v_0}{\lam} \rho + \frac{\mu^\text{c}}{\lam}F_0 p_x \right) \nonumber \\
  &= L_y \left[ \frac{C_-^{(1)} \left(v_0 + \mu^\text{c} F_0 / \sqrt{2}\right)}{\beta F_0 + v_0/\left(\sqrt{2}D_0\right)} + \frac{C_-^{(2)} \left(v_0 - \mu^\text{c} F_0 / \sqrt{2}\right)}{\beta F_0 - v_0/\left(\sqrt{2}D_0\right)}  \right]
  \label{app_eq:flux_minus}
\end{align}
for subsystem $\Omega_-$. 

\section{Simulation details}

\subsection{Explicit model}
\label{app:sim_explicit}

We simulate an $L\times L \times L$ box with $L = 8 r_\text{C}$ where $r_\text{C}$ is the radius of the catalyst at the origin. We model the solute-catalyst interactions as excluded volume via the repulsive short-ranged Weeks-Chandler-Anderson potential
\begin{equation}
    u(r; \epsilon, \sig) = \begin{cases}
    4\epsilon \left[ \left(  \frac{\sig}{r}\right)^{12} - \left( \frac{\sig}{r}\right)^6   \right] + \epsilon &\text{if } r < 2^{1/6}\sig  
    \\
    0 &\text{otherwise}
  \end{cases}
  \label{app_eq:wca}
\end{equation}
with $u_\al(r) \equiv u(r; \epsilon_\al, r_\text{C})$. We employ $\epsilon_\text{S} = 10$ for substrate and $\epsilon_\text{P} = 1$ for product solutes. Inside the catalytic zone, we attempt a chemical event with probability $p_0 = 0.1$ each time-step, fixing an attempt rate $k_0 = p_0 / \delta t$ where $\delta t = 10^{-4}$ is the discretisation of time (in Brownian units) for integration of the translational degrees of freedom.

The repulsive potential of the catalyst shifts the equilibrium concentrations by decreasing the effective system volume. The solute molecule population is consequently asymmetric even in a state of thermal equilibrium [$N^\text{eq}_\text{S} \neq N^\text{eq}_\text{P} \neq (N_\text{S} + N_\text{P})/2$] and shifts the chemical potential difference by the (logarithm) of equilibrium constant $K = N^\text{eq}_\text{S}/ N^\text{eq}_\text{P}$. From the partition function
\begin{align}
    Z(N_\text{S}, N_\text{P}) = \frac{\nu^{N_\text{S}}_\text{S}\nu^{N_\text{P}}_\text{P}}{\lam_T^{3(N_\text{S} + N_\text{P})}N_\text{S}!N_\text{P}!},
\end{align}
with thermal de Broglie wavelength $\lam_T$ and effective volumes 
\begin{align}
    \nu_\al \equiv V + 4\pi \IInt{r}{0}{r_\text{C}} r^2 \left( e^{-\beta u_\al(r)} -1 \right)
\end{align}
we find $K = \nu_\text{S}/\nu_\text{P}$ upon setting the chemical potentials equal. For small packing fractions of catalytic particles, i.e., large system volume $V$, their effect on the equilibrium population is negligible as $K \to 1$, leaving equal fractions of substrate and product molecules. The constant $K$ introduces an effective ideal contribution to the chemical potential, that we inserted into Eq.~\eqref{eq:explicit-chemostat} in the main text in order to use the relative concentrations in the reservoir $\Delta \mu = \ln{\bar{c}_\text{S} / \bar{c}_\text{P}}$ as the control parameter for activity.

\subsection{Active Brownian particles}
\label{app:sim_MIPS}

To investigate the dissipation rate of particles undergoing MIPS we simulate a suspension of $N=4000$ active particles of diameter $\sig$ in a rectangular $L\times L$ box, with edge length $L \approx 102.3$ and periodic boundary conditions. The global packing fraction is $\bar \phi =0.3$, corresponding to a density $\bar \rho \approx 0.38$. Particles are self-propelled with speed $v_0 = 80$, which results in a drive amplitude of $\Delta \mu \approx 2.2$, by setting $\lam = 10^{-3}$ and $\keff\lam=50$. Note that we employ dimensionless quantities by reporting all numerical values of lengths in units of $\sig$, energies in units of $\kT$, and times in units of $\sig^2/D_0$. The rotational diffusion coefficient is set to $D_\text{r} = 3$. Assuming particles to be hard Brownian disks, we integrate their equations of motion 
\begin{align}
  \dot{\vec r}_i = v_0 \vec e_i - \mu_0 \nabla_i U + \sqrt{2D_0}\,\nois_i, \qquad \dot \vhi_i = \sqrt{2D_\text{r}}\,\xi^\text{r}_{i} 
\end{align}
with time step $\delta t = 5\times 10^{-6}$. Here components of $\nois_i$ and $\xi^\text{r}_{i}$ are drawn from a uniform distribution over $\left[-\sqrt{3}, \sqrt{3}\right]$. We model excluded volume via the repulsive short-ranged Weeks-Chandler-Anderson potential $u(r; \epsilon_0, \sig)$ [see Eq.~\eqref{app_eq:wca}], with potential energy well given by $\epsilon_0=100$ and $r$ the distance between colloidal particles. The potential is $U = \sum_{i<j} u(|\vec r_{i} - \vec r_{j}|)$.

The data presented in Fig.~\ref{fig:MIPS}(b-d) is obtained by dividing the simulation box into equally sized bins and calculating concerned averages within each such bin. To improve statistics, discrete binned data is time averaged over several snapshots with sampling frequency $\tau^\prime = 0.5$. Spatial derivatives entering the dissipation rate [see Eq.~\eqref{eq:lepr:fields}] are calculated by employing a central difference scheme, respecting periodic boundary conditions, on the discretized data.

%% ---- bibliography ----

%merlin.mbs apsrev4-1.bst 2010-07-25 4.21a (PWD, AO, DPC) hacked
%Control: key (0)
%Control: author (8) initials jnrlst
%Control: editor formatted (1) identically to author
%Control: production of article title (0) allowed
%Control: page (1) range
%Control: year (1) truncated
%Control: production of eprint (0) enabled
%

\end{document}